# The Balloon-borne Investigation of Temperature and Speed of Electrons in the Corona (BITSE): Mission Description and Preliminary Results

N. Gopalswamy*, J. Newmark, S. Yashiro[1], P. Mäkelä[1], N. Reginald[1],

N. Thakur[1], Q. Gong

*NASA Goddard Space Flight Center, Greenbelt, MD*

Y-H. Kim, K-S. Cho[2], S-H. Choi, J-H. Baek, S-C. Bong, H-S. Yang, J-Y. Park, J-H. Kim, Y-D. Park, J.-O. Lee, R.-S. Kim[2], E.-K. Lim

*Korea Astronomy and Space Science Institute, Daejeon City, Republic of Korea*

[1]*Also at The Catholic University of America, Washington, DC*

[2]*University of Science and Technology, Daejeon, Korea*

*Corresponding author. E-mail: nat.gopalswamy@nasa.gov

Abstract

We report on the Balloon-borne Investigation of Temperature and Speed of Electrons in the corona (BITSE) mission launched recently to observe the solar corona from ≈3 Rs to 15 Rs at four wavelengths (393.5, 405.0, 398.7, and 423.4 nm). The BITSE instrument is an externally occulted single stage coronagraph developed at NASA's Goddard Space Flight Center in collaboration with the Korea Astronomy and Space Science Institute (KASI). BITSE used a polarization camera that provided polarization and total brightness images of size 1024 × 1024 pixels. The Wallops Arc Second Pointing (WASP) system developed at NASA's Wallops Flight Facility (WFF) was used for Sun-pointing. The coronagraph and WASP were mounted on a gondola provided by WFF and launched from the Fort Sumner, New Mexico station of Columbia Scientific Balloon Facility (CSBF) on September 18, 2019. BITSE obtained 17,060 coronal images at a float altitude of ≈128,000 feet (≈39 km) over a period of ≈4 hrs. BITSE flight software was based on NASA's core Flight System, which was designed to help develop flight quality software. We used EVTM (Ethernet Via Telemetry) to download science data during operations; all images were stored on board using flash storage. At the end of the mission, all data were recovered and analyzed. Preliminary analysis shows that BITSE imaged the solar minimum corona with the equatorial streamers on the east and west limbs. The narrow streamers observed by BITSE are in good agreement with the geometric properties obtained by the Solar and Heliospheric Observatory (SOHO) coronagraphs in the overlapping physical domain. In spite of the small signal-to-noise ratio (≈14) we were able to obtain the temperature and flow speed of the western steamer. In the heliocentric distance range 4 – 7 Rs on the western streamer, we obtained a temperature of ≈1.0±0.3 MK and a flow speed of ≈260 km s$^{-1}$ with a large uncertainty interval.

Keywords:

Coronagraphs; solar wind; solar wind density; solar wind temperature; solar wind speed; coronal spectrum

# 1. Introduction

The Balloon-borne Investigation of Temperature and Speed of Electrons in the corona (BITSE) mission's primary instrument is an optical telescope fitted with an external occulter that blocks the direct sunlight so the faint emission from the corona can be observed. Unlike traditional coronagraphs, the BITSE coronagraph (BITSE COR) has a single-stage optics and a polarization detector to obtain both the total and polarized brightness of the solar corona at four narrow passbands in the blue end of the K-coronal spectrum. The BITSE mission is specifically built to demonstrate that the temperature and flow speed of the coronal electrons can be measured by quantifying the change in shape and red shift of the K-corona spectrum when the temperature and speed of the electrons change (Menzel and Pasachoff, 1968; Cram, 1976; Ichimoto et al., 1996;1997; Takahashi, Yoneshima, and Hiei, 2000; Reginald and Davila, 2000; Reginald et al., 2003; 2011). The temperature and flow speed of the electrons along with the density are key input parameters to models of solar wind acceleration. Traditionally, coronagraphs measure just the electron density, while the BITSE coronagraph utilizes spectral information to determine the electron temperature and flow velocity in addition to density.

Direct measurements of electron temperature have not been generally made beyond ≈3 solar radii (Rs) until the domain of in situ measurements. There is only one temperature value (1.1 ±0.3 MK) at a distance of 2.7 Rs obtained by Fineschi et al. (1998) from the measurement of the electron-scattered Lyα profiles in the corona. Electron temperature measurements in the 3 – 8 Rs distance range are necessary to determine the bulk-plasma heating rate in different solar wind regions, as well as the energy partitioning between protons and electrons (see e.g., Cranmer et al., 2010). The energy partitioning is a key diagnostic of turbulence models (e.g., Cranmer, 2012; Cranmer and van Ballegooijen, 2012) as well as a driver of the stability of helmet streamers (e.g., Endeve, Holzer, and Leer, 2004). Electron temperature measurements are also important in understanding the sunward conduction of electron thermal energy that determines the solar wind mass flux (Withbroe, 1988; Lie-Svendsen et al., 2002). Solar wind speed measurements are also rare in the solar wind acceleration region (see Abbo et al., 2016). Grall et al. (1996) used the Very Long Baseline Array (VLBA) to observe interplanetary scintillation of distant radio sources viewed through the corona to show that the fast wind speed in the polar region is ≈600 km s$^{-1}$ at a distance of ≈6.8 Rs indicating that most of the solar wind acceleration has occurred already. Strachan et al. (2002) combined the data from the Ultraviolet Coronagraph Spectrometer (UVCS) and the Large Angle and Spectrometric Coronagraph (LASCO) on board the Solar and Heliospheric Observatory (SOHO) to show that there is no measurable O5+ outflow along the axis of an equatorial streamer until after ≈3.5 Rs, followed by a rapid increase in velocity to ≈90 km s$^{-1}$ at 5 Rs in the streamer stalk. Dolei, Spadaro, and Ventura (2015) combined UV spectral and white-light observations to derive a flow speed in the range 40 – 140 km s$^{-1}$ at distances of 2.5 – 5.0 Rs at the edges of streamers.

Thus, there is a clear need for measurements of electron temperature and flow speed in the solar wind acceleration region, 3 – 8 Rs for constraining solar wind acceleration and heating. The measurement gap is bracketed by spectroscopic measurements made at distances < 3 Rs and the Parker Solar Probe measurements to be at distances > 9 Rs. For example, the state-of-the-art solar wind models assume an electron temperature variation to obtain the flow speed. Measured electron temperature input to the models will make the models more realistic. BITSE demonstrates that the electron temperature and flow speed can be measured, providing critical observational constraints.

The BITSE mission is the culmination of several technological activities in coronagraphy undertaken at NASA's Goddard Space Flight Center (NASA/GSFC) over the past several years (Gopalswamy and Gong, 2018): (i) A breadboard version of a single-stage, externally occulted coronagraph known as the Goddard Miniature Coronagraph (GMC) was built and tested in the laboratory and in vacuum tank, and (ii) a commercial polarization camera was tested during the August 21, 2017 total solar eclipse in the United States, that obtained both polarization and total brightness images in four narrow band filters (Reginald et al., 2017; Gopalswamy et al., 2018; Cho et al., 2020). The polarization camera allows us to image the corona at different polarization positions simultaneously, instead of sequentially observing at different polarization positions using a polarization wheel. Eclipses last typically only for a couple of minutes, barely long enough obtain images in all filters. A balloon platform offers a much longer period of observations. Balloon-borne coronagraphs observed at long wavelengths have provided valuable information on the density distribution of the corona (see Koutchmy, 1988, for a review of past balloon-borne coronagraphs). The BITSE mission demonstrated both the GMC and the polarization camera for the first time in space.

It must be noted that Cram's (1976) technique has been attempted during total solar eclipses in the past. Ichimoto et al. (1996) obtained coronal spectra in the range 3700 to 4729 Å at two heights (1.1 and 2 Rs) during the November 3, 1994 total solar eclipse, compared them with the expected spectra for a given coronal electron density distribution, and obtained an electron temperature of 1.6 MK in a streamer. They also estimated the bulk flow speed of the electrons as ≈80 km s$^{-1}$ from the change in the intensity ratios at 3990 and 4218 Å. Takahashi, Yoneshima, and Hiei (2000) used the intensity ratios 4230 Å/4100 Å and 4220 Å/4100 Å to show that the streamers were hotter than the coronal holes, although they did not report actual temperatures. Reginald and Davila (2000) and Reginald et al. (2003) used the Multiaperture Coronal Spectrometer (MACS) to measure white-light K-coronal intensities simultaneously at several locations in the corona during the total solar eclipse of June 21, 2001. They obtained an electron temperatures of 0.96 ± 0.05 and 1.2 ± 0.2 MK in two streamers at a heliocentric distance of 1.1 Rs. The corresponding flow speeds were 72 km s$^{-1}$ and 257 km s$^{-1}$ with large uncertainties. Eclipse observations have thus demonstrated the Cram's technique, but the signal-to-noise ratio was not enough beyond 1.5 Rs to obtain the physical parameters in the solar wind acceleration region.

In this paper we present an overview of the BITSE mission including the instrument, pointing system, on board computer, and the balloon launch. We also describe various subsystems: the BITSE electronics box (BEB), BITSE camera, filter wheel, optical bench, coronagraph door, thermal control, and the power distribution unit (PDU). Finally, we present images obtained by BITSE and preliminary results on the temperature and flow speed of electrons.

## 2. The Coronagraph

BITSE is a single-instrument mission. The instrument, BITSE COR is an externally occulted single-stage coronagraph that images the solar corona at the blue end of the solar spectrum. Single stage means that the optical system does not have an internal occulter and a Lyot stop employed in traditional multi-stage coronagraphs. The coronagraph images the photospheric light Thomson-scattered by coronal electrons that may have different temperatures and flow speeds in different coronal magnetic structures. The coronagraph consists of an optical assembly, an external occulter, straylight baffles, a heat rejection mirror, a filter wheel, and a polarization camera. The coronagraph has only two flight mechanisms: aperture door (one-time opening) and a filter-wheel rotator. This represents one less mechanism (no polarizer wheel) compared to the traditional coronagraphs. The coronagraph is pointed at the Sun using the GSFC/Wallops Arc-Second Pointer (WASP) system. The polarization camera obtained the images that were stored on board and a fraction of the data were sent in real time to the ground station.

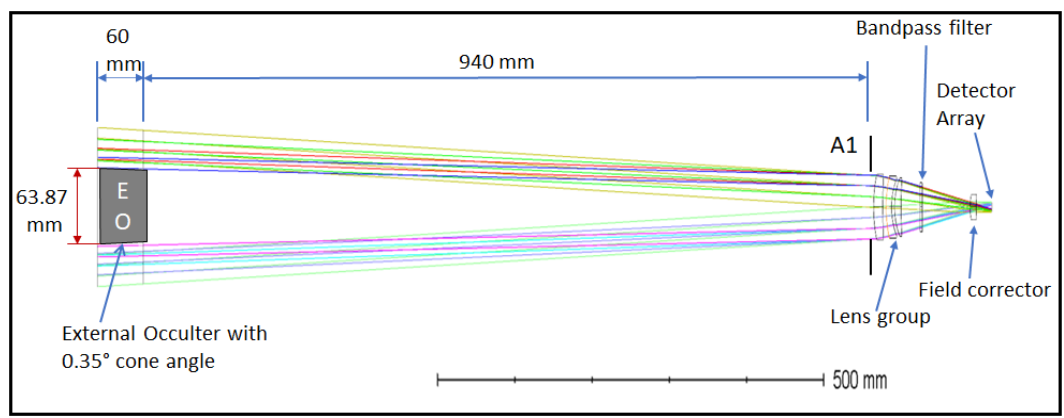

Figure 1. Optical layout of BITSE COR. Sunlight enters the coronagraph from the left. The external occulter (EO) is a right frustum with a front diameter of 63.87 mm and has a cone angle of 0.35º. A1 aperture has a diameter of 50 mm. The 1-m separation between the A1 aperture and EO is to reduce vignetting. Coronal light is focused by the lens group and imaged at the detector array after passing through the bandpass filter and the field corrector lens.

**Table 1.** The design parameters of the BITSE COR optical assembly

| Parameter | Value |
|---|---|
| Outer FOV cutoff (from Sun center) | 15 $R_S$ |
| Inner FOV cutoff (from Sun center) | 3 $R_S$ |
| Wavelength (λ) range | 380 – 460 nm |
| Effective focal length | 103 mm |
| Entrance pupil diameter (A1) | 50 mm |
| Detector array[a] | CCD, 2048 × 2048, 7.4 µm pixel |

[a]The number of pixels in the CCD is 2128 × 2112, the effective number of pixels is 2080 × 2080, the number of active pixels is 2048 × 2048.

## 2.1 Optical design

Figure 1 shows the optical layout of BITSE COR with the sunlight entering the system from the left. Table 1 lists the design parameters of the coronagraph. The detector covers a heliocentric radius of ≈ 15 Rs, however, the image quality is optimized for < 8 Rs, covering the key region of interest for temperature and speed measurements. The external occulter (EO) is a 60-mm long right frustum with a front diameter of 63.87 mm and a cone angle of 0.35º. EO blocks the photosphere and the inner corona at a heliocentric radius of 1.5 Rs, which is smaller than the specified inner field of view (FOV) cutoff at 3.0 Rs. The inner cutoff is determined by a circular occulter mask attached to the detector array (Gong, Gopalswamy, and Newmark, 2019). The occulter mask on the face of the detector has a diameter of 2.5 mm that corresponds to an inner FOV cutoff at 3 Rs. Based on simulations, it was found that the diffraction straylight level beyond 3 Rs is not sensitive to the EO cutoff from 1.5 Rs to 3 Rs. The only case with higher level of diffracted light is when EO blocks just the photosphere (1 Rs). We chose 1.5 Rs because it decreases the vignetting significantly near the specified inner cutoff (3 Rs) without compromising the diffraction suppression. The long separation between EO and the A1 aperture reduces the beam vignetting and increases the signal-to-noise ratio to compensate for the noise caused by the sky brightness at the float altitude of 128,000 ft (≈39 km). Table 2 shows the specifications of optical elements used in the coronagraph. L1 – L3 constitute the lens group with front (S1) and back (S2) surfaces having different shapes and radii of curvature. All optical filters are flat. The lenses L1, L2, and the field corrector (FC) have the convex-concave shape. Both surfaces of L3 are convex.

A heat rejection mirror (HRM) is located in front of the lens group and the filters are inserted after the lens group. HRM images the sunlight in the gap between the EO and the telescope-tube baffles and reflects it back to the sky.

**Table 2.** Coronagraph optical elements and their specifications

| Element | Radius of Curv. (mm) | Thickness (mm) | Material | Diameter (mm) | Shape |
|---|---|---|---|---|---|
| L1/S1 | 80.912 | 12.0 | Fused silica | 56.0 | Convex |
| L1/S2 | 336.561 | | | | Concave |
| L2/S1 | 74.883 | 7.0 | PBM18Y | 52.0 | Convex |
| L2/S2 | 40.558 | | | | Concave |
| L3/S1 | 41.931 | 10.4 | $CaF_2$ | 50.0 | Convex |
| L3/S2 | 251.659 | | | | Convex |
| Filter/S1 | ∞ | 2.0 | S-BSL7 | 47.0 | Flat |
| Filter/S2 | ∞ | | | | Flat |
| FC/S1 | 30.232 | 8.212 | CaF2 | 20.0 | Convex |
| FC/S2 | 146.683 | | | | Concave |

**2.2 Mechanical design**

The BITSE COR mechanical design consists of the following primary components: a modular aluminum structure for the telescope tube, external occulter, aperture door mechanism, lens group assembly, filter wheel, polarization camera, thermal radiator system, heat rejection mirror, base plate, the BITSE electronics box (BEB), and the power distribution unit (PDU) (see Fig. 2). The main coronagraph tube is 1004 mm long with a diameter of 260 mm. The external occulter is located at the front edge of the coronagraph tube. BITSE COR also has a 607 mm long and 343 mm diameter straylight baffle tube to prevent glint from the balloon/gondola or earthshine enter the coronagraph. Both tubes are made of a series of conventionally machined aluminum tube sections that are bolted and pinned together for alignment stability. To reduce stray light, each tube section is black anodized, and the front tube section has a series of integral knife-edge baffles. The stray-light baffle tube has five baffles with progressively decreasing clear aperture sizes except for the first baffle which has the smallest clear aperture 244.8 mm in diameter. The design of the stray-light baffle is based on forward and reverse raytracing that demonstrated that no light ray making an angle > 5º from the optical axis enters the tube. The sixth baffle is in the occulter plane (A0) with a sawtooth inner edge. The tooth design is such that the diffraction normal of the edge is located clear of the objective lens inside the A1 aperture. Baffles 7 – 16 are in the coronagraph tube, with progressively increasing clear aperture sizes from 202.93 mm at Baffle 7 to 210.39 mm at Baffle 15 except for the last baffle. The last baffle (16) is at the A1 aperture and has a clear aperture of 198 mm in diameter. The coronagraph tube also has a purge gas inlet to keep the optical path clean. The aperture door mechanism consists of a spring drive with a resettable pin puller, a doorstop to dissipate the door's kinetic energy, and a commercial controller. The drive electronics of the controller is enclosed in a custom enclosure and mounted on the optical bench. The pin puller mechanism is mounted on top of the front baffle tube, while the doorstop is mounted on the bottom side of the tube (see Fig.2).

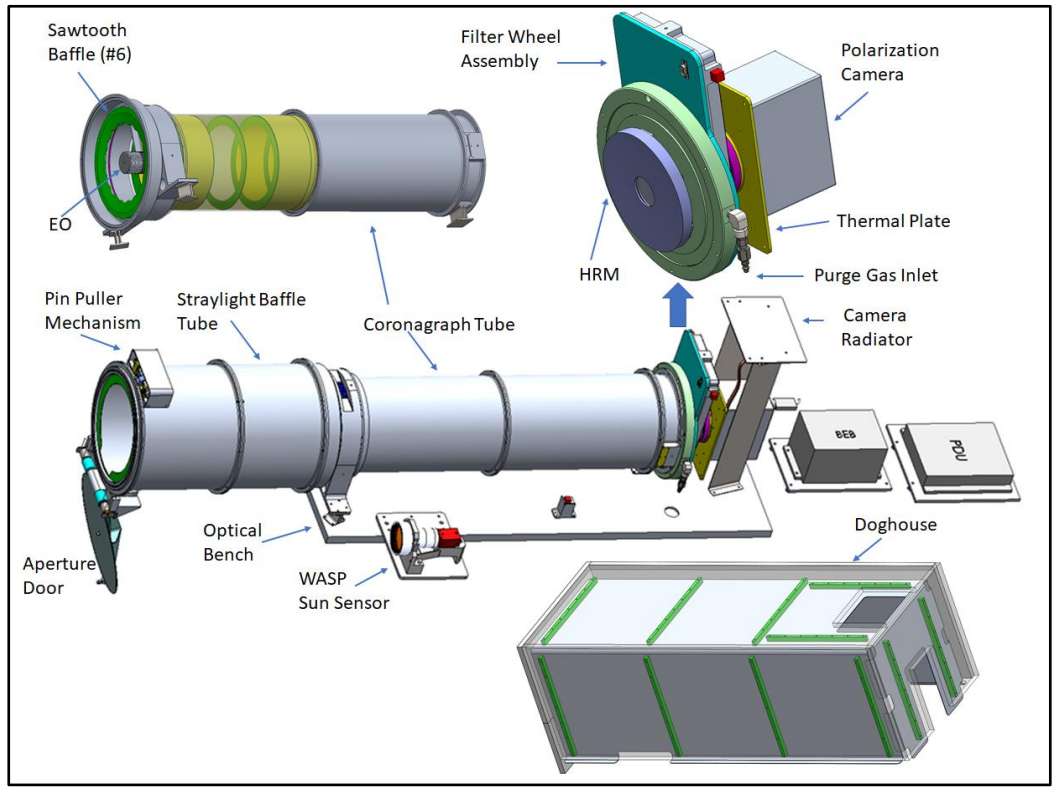

**Figure 2.** Mechanical design of BITSE COR along with the BITSE electronics box (BEB) and the power distribution unit (PDU). The bulkhead assembly consisting of the heat rejection mirror (HRM), the lens group assembly, the filter wheel assembly, and the polarization camera are shown separately. A few of the baffles inside the coronagraph tube are also shown separately. A thermal enclosure ("doghouse") covers the coronagraph over the base plate from the polarization camera at the backend to the front end of the coronagraph tube. It has three cutouts for the camera radiator, harnesses, and the front baffle. The straylight baffle tube is not covered by the thermal enclosure. The door mechanism is for opening once when the instrument reaches float altitude. The instrument total mass is 116 kg.

The bulkhead assembly consists of an HRM, a lens-group subassembly and a filter wheel (see Fig. 2). HRM reflects the solar energy out through the front aperture and directs it away from the occulter assembly and the front tube's internal surfaces. The lens group consists of 3 lenses and 3 lens cells. Each lens is aligned and bonded into a cell and each cell is aligned to the adjacent cell and then bolted into its aligned position (see Fig. 2). Each lens is positioned onto to a precision flat surface in each cell for tip/tilt alignment. Precision dowel pins are used to center the lens in the cell then a room-temperature-vulcanizing (RTV) adhesive is injected into the cell to bond the lens to the cell; the pins are removed after the adhesive cures. The filter wheel drive mechanism is supported by a pair of angular contact ball bearings and a piezo-motor is operated in an open-loop to position the filter elements into the chief ray. A hard stop is used to reposition the wheel to a home position if motor step counts are lost. The camera is mounted onto the aft section of the assembly and a thermal radiator is used to cool the camera's CCD. The camera enclosure is placed inside a 1-atm pressure vessel because the camera is not vacuum compatible. A base plate (optical bench) supports the instrument assembly, the Sun sensor, and a thermal enclosure. Also mounted on the base plate are the aperture-door controller box and the master reference cube for alignment. The base plate also has a hole providing access to the purge hose that connects to the gas inlet in the bulkhead assembly. The base plate, the BITSE electronics box (BEB), and the power distribution unit (PDU) are mounted on the 10-inch telescopic beam, which is part of the WASP system mounted on the balloon gondola.

## 2.3 Subsystems
### 2.3.1 Filter Wheel Assembly

The filter wheel is a key component of the passband ratio imaging technique. Figure 3 shows the filter wheel assembly (FWA) with its three elements: the frame with five slots, power converter, and motion controller. The wheel diameter is 156.4 mm. The filter positions are at a distance of 49 mm from the shaft. The filter slots have a diameter of 51.5 mm. It typically takes 1 s to move and position a filter in the optical path of the coronagraph (the move corresponds to an angular rotation of 72º). The motion controller has a piezo motor and an optical encoder with a positioning reliability better than 0.5 mm. A three-point pivot alignment system for each filter and bearings around the filter wheel are used to achieve a tilt reliability of ≈ 9 arc min. A thermal/vacuum test of the filter wheel confirmed that it can operate in the temperature range -10ºC to +40ºC. The FWA power consumption is 2 W (normal) and 4 W (active).

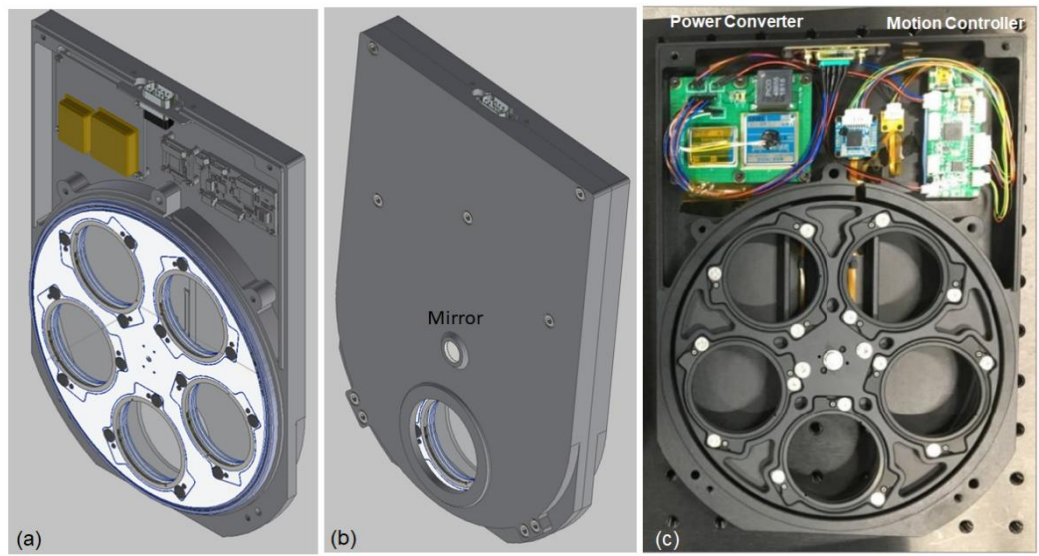

Figure 3. The filter wheel assembly. (a) solid model without the cover, (b) solid model with the cover closed and alignment mirror, and (c) picture of a test version of the filter wheel (without cover). The power converter and motion controller boards can be seen at the top with a D-Sub 9 male port. The voltage input to the FWA is 30.8 V unregulated. A – b

**Table 3** Filter specifications

| Filter # | Central λ (nm) | Bandwidth (nm) | Purpose |
|---|---|---|---|
| 1 | 393.5 ± 0.25 | 6 ± 0.25 | Temperature |
| 2 | 405.0 ± 0.25 | 5 ± 0.25 | Temperature |
| 3 | 398.7 ± 0.25 | 6 ± 0.25 | Speed |
| 4 | 423.4 ± 0.25 | 5 ± 0.25 | Speed |
| 5 | 405.0 ± 1 (380 – 430) | 50 ± 1 | density |

Table 3 shows the specifications of the four narrowband filters and the broadband filter. The central wavelengths are slightly different from the previously employed wavelengths (e.g., in the recent total solar eclipse set up). The substrate of the filters is Ohara S-BSL7 optical glass (equivalent to BK7) with a diameter of 47 mm and thickness of 2 mm. The filters have an antireflection (AR) coating on the back surface. The filter transmittance is over 90% average in the defined bandwidth. The transmittance drops from 90% of the maximum to 10% within 20% of the bandwidth. The filters have an out-of-band rejection to the level of $1\times10^{-3}$ from 300 nm to 1100 nm. The filters are optimized for an incident angle of 8º cone for central wavelength.

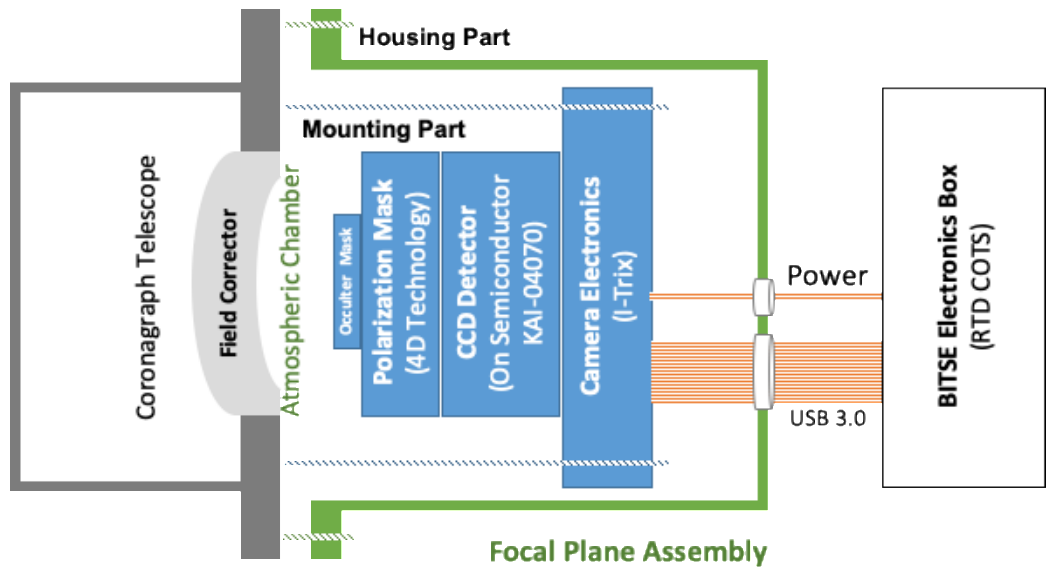

**Figure 4.** Block diagram of the focal plane assembly. The field corrector is at the front wall of the pressure chamber and the power and data connection (USB 3.0) ports are at the backend of the camera housing.

**2.3.2 Focal Plane Assembly**
The Focal Plane Assembly (FPA) consists of a polarization camera built from commercial parts. The polarization camera is similar to the one used during the 2017 total solar eclipse observation but uses a new CCD sensor KAI-04070 and custom-made electronics. A micropolarizer array (MPA) is glued to the front face of the CCD. The focal plane occulter mask is bonded to the central part of the MPA. All camera parts are mounted inside a housing with known thermal dissipation characteristics. The polarization camera is then enclosed inside a 1-atm pressure chamber. The field corrector is at the front wall of the pressure chamber. The atmospheric chamber keeping the 1 atm environment protects the MPA and occulting disk bonding.

Figure 4 shows a schematic of the FPA at the end of the coronagraph telescope. The FPA sensor is an On Semiconductor Interline CCD chip KAI-04070 with $2048 \times 2048$ active pixels, the physical size of a pixel being 7.4 µm. The active image area is 15.2 mm × 15.2 mm. The quantum efficiency (QE) of the CCD ranges from 40% to 52% in the range 380 – 455 nm that includes BITSE operating wavelengths. The lower QE corresponds to shorter wavelengths. The MPA with the occulter mask is bonded to the front face of the CCD and provides $1950 \times 1950$ pixels for polarization measurements. The occulter mask is a black stainless-steel metal disk with a diameter of 2.5 mm and thickness of 25 µm bonded to the geometrical center of the MPA front surface. The MPA's substrate is 0.7 mm thick and made of fused silica with UV AR coating on the backside. The camera electronics boards (schematic and pictures) are shown in Fig. 5 along with the assembled camera. The camera enclosure has three sections. The front section has the face plate with a slot in which the CCD sensor is mounted. The front plate also serves as a reference position. The sensor board is attached to the backside of the plate followed by a support plate. The middle section of the enclosure holds the analog front-end (AFE) board on the front and the image capture board on the rear. The rear section of the enclosure holds the interface board on the frontside and has the back plate on the rear side. The back plate has two ports for data (USB3.0, A-type female connector) and power (4-pin circular connector), and a circular mirror (diameter 12.7 mm) for alignment. The three sections are brought together to form the whole camera. Thermal plates are placed on the inside of the enclosure that touch the heat sources in the boards with a thermal pad.

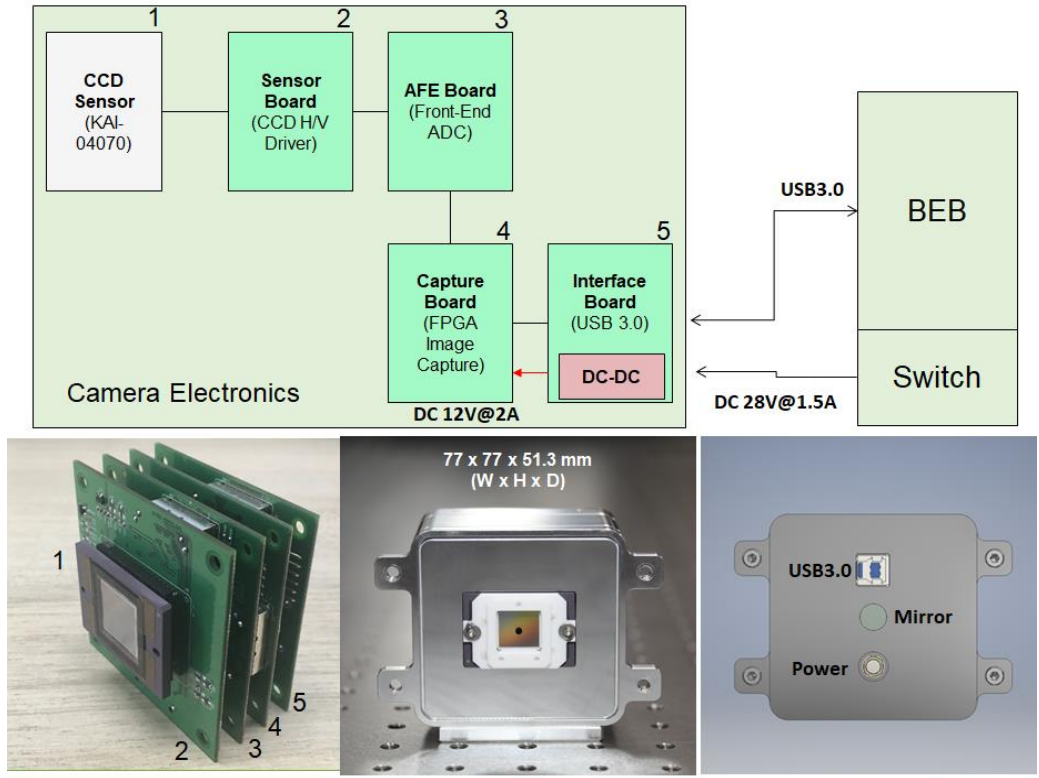

**Figure 5.** (top) Block diagram of the camera electronics with the four electronic boards and the connection to BEB and power. CCD sensor (1) is attached to the sensor board (2), followed by the Analog Front-End (AFE) board (3) for the analog-to-digital converter (ADC). The image capture board (4) is for the field-programmable gate array (FPGA) device (Xilinx ARTIX 7). The interface board (5) is for the USB. (bottom left) picture of the sensor and the four boards in the camera. (bottom middle) Front view of the camera (Sun-facing side) with the black metal disk on the MPA (black dot with a diameter of 2.5 mm). The CCD sensor is mounted on the front plate of the enclosure, which serves as a reference position. The camera enclosure dimensions (width × height × depth) are: 77 × 77 × 51.3 mm. (bottom right) rear face of the camera enclosure with ports for data (USB3.0 A-type female connector) and power (4-pin circular connector). The circular mirror of diameter 12.7 mm is for alignment purposes.

**3 BITSE Electronics Box**

BITSE Electronics Box (BEB) is commercial RTD box, centered on an Intel Core i7 Dual-Core single board computer and is central to the coronagraph operation as detailed in Fig. 6. BEB functions are storing science and housekeeping data, data communication interface for subsystem control and data handling, software running environment, and power relay. BEB and flight software are designed for automatic operation. The modular packaging of IDAN (Intelligent Data Acquisition Node) is used to stack the boards in BEB. The rugged modular enclosure serves as a passive heat sink and is made from solid aluminum alloy and coated with aluminum chromate. The physical size of the enclosure (width × height × depth) is 151 mm × 130 mm × 213 mm with a mass of 6.4 kg including customization heat sinks. The BEB finishes with teflon-silver tapes. The BEB operating temperature range is -40 to +85ºC. All the boards in the BEB have been tested for vacuum compatibility.

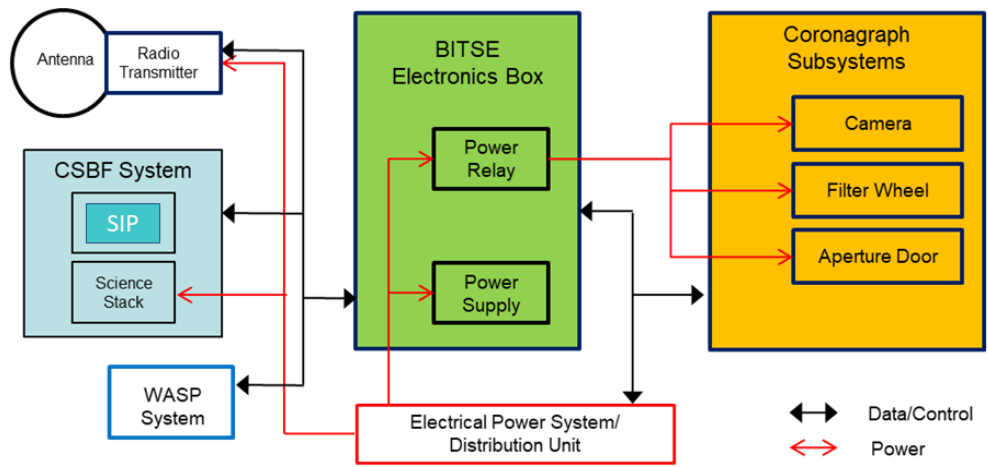

**Figure 6.** A block diagram showing how the BITSE electronics box (BEB) is central to the BITSE hardware. BEB controls all coronagraph subsystems and communicates with the radio transmitter, WASP system and CSBF system. The WASP system includes the pointing system and GPS. Part of the Support Instrumentation Package (SIP) is the Ethernet-Via-Telemetry (EVTM) system for science data downlink (8 Mbps). The science stack provides analog and digital return telemetry and discrete command outputs.

Figure 7 shows the picture of the BEB taken during a laboratory test. The modular nature is indicated by the lines on the top face of the box. Two of the seven BEB boards are also shown in Fig. 7: the single board computer (SBC) and the power supply. The SBC provides software operation environment, accepts a power reset signal from the support instrument package (SIP), interfaces with WASP and filter wheel, and acquires aperture door status. The power supply provides power to peripheral component interconnect express (PCIe) stackable modules. The other boards are: power relay (to relay power to FWA, camera, and aperture-door power control), analog/digital data acquisition (to interface with the PDU and aperture door), serial port (for interfacing with mini-SIP and radio transmitter), USB3 controller (to interface with the polarization camera), and data storage (for housekeeping and science data archival in a 512 GB solid-state drive). The boards are mounted independently on their own modular frames and all I/O connections are accessible on the outside of the frame as seen in Fig. 7.

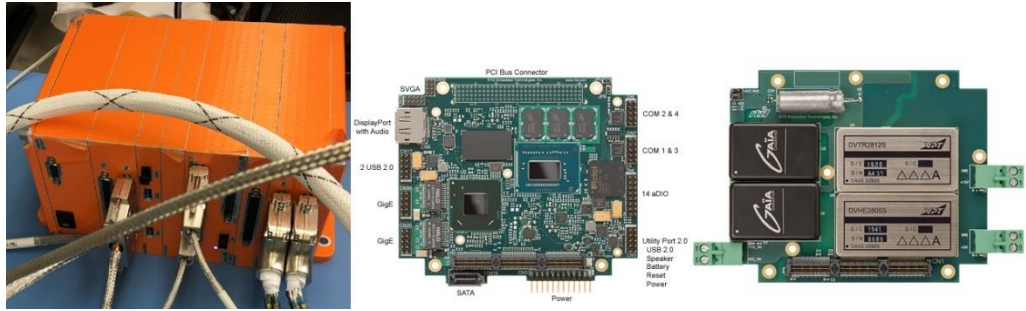

**Figure 7.** (left) The BITSE electronics box with connectors for a laboratory test. The orange-colored skin of the BEB is a removable protection film on Teflon-silver tapes. (middle) The single board computer (SBC) that uses Intel® Core™ i7 3517UE (Dual-Core, 1.7 to 2.8GHz, error-correcting code (ECC) Memory supported) and includes a 4GB DDR3 SDRAM (ECC), and a 32GB on board (ECC) solid state drive. (right) Highly reliable avionics power supply board with an output power of 40 W at 5 V (8 amps DC) and at 12 V (3.33 amps DC). The accepted input voltage range is 16 – 40 V DC with a maximum input power of 100.64 W.

**4 Thermal Design**

The BITSE thermal design is based on the following considerations: (i) the operating temperature at the camera interface plate is 7⁰C to 8⁰C so that the CCD/mask runs at ≈15˚C, (ii) BITSE is launched with the coronagraph powered on and no survival heaters are provided, (iii) BITSE needs to be thermally stable when the instrument reaches the float altitude. One of the important design considerations is the fact that the instrument has to transit through the tropopause (altitude range 15 – 25 km) where the air temperature reaches -59⁰C.

The thermal control system employs a thermal design utilizing passive radiators, heaters, thermal straps, and insulation (Styrofoam and multilayer insulation, MLI). The telescope assembly (excluding the BEB, PDU, and the front baffle) is shrouded within a doghouse structure maintained at room temperature via thermostatically controlled heaters. The front baffle tube is thermally isolated from the telescope and enclosed in Styrofoam insulation. The optical bench is thermally isolated from the gondola telescope beam. The BEB and PDU are mounted to individually controlled heater baseplates that are attached to the telescope beam. The camera interface plate is software controlled to stay at 7⁰±0.1⁰C such that the CCD and mask are maintained at 15˚C. A passive radiator is used to cold bias the camera and two Kapton film heaters are located on the plate to provide positive temperature control. The camera assembly is covered with MLI and thermally isolated from the telescope to minimize parasitic heat transfer. The camera radiator is mounted to, and thermally isolated from, the optics.

Figure 8 shows the thermal model of BITSE divided into eight zones for thermal control and the placement of heaters in each zone. Heaters with different dimensions are used depending on the shape of the thermal zones. Heaters are sized based on a minimum voltage of 29.15V provided. Heaters for each circuit are wired in parallel. Each circuit (except for the camera heater) has 2 thermostats for redundancy. Thermostats are wired in series (they are typically fail-closed). The camera heater is software controlled due to tighter temperature range (6.9ºC to 7.1ºC). The temperature range of all the other zones is 18ºC to 22ºC, except for the BEB and PDU base plates, which have a range of -2ºC to 2ºC. The operational power has been estimated as 133, 64, and 235 W for the float-altitude, pre-launch, and ascent phases, respectively.

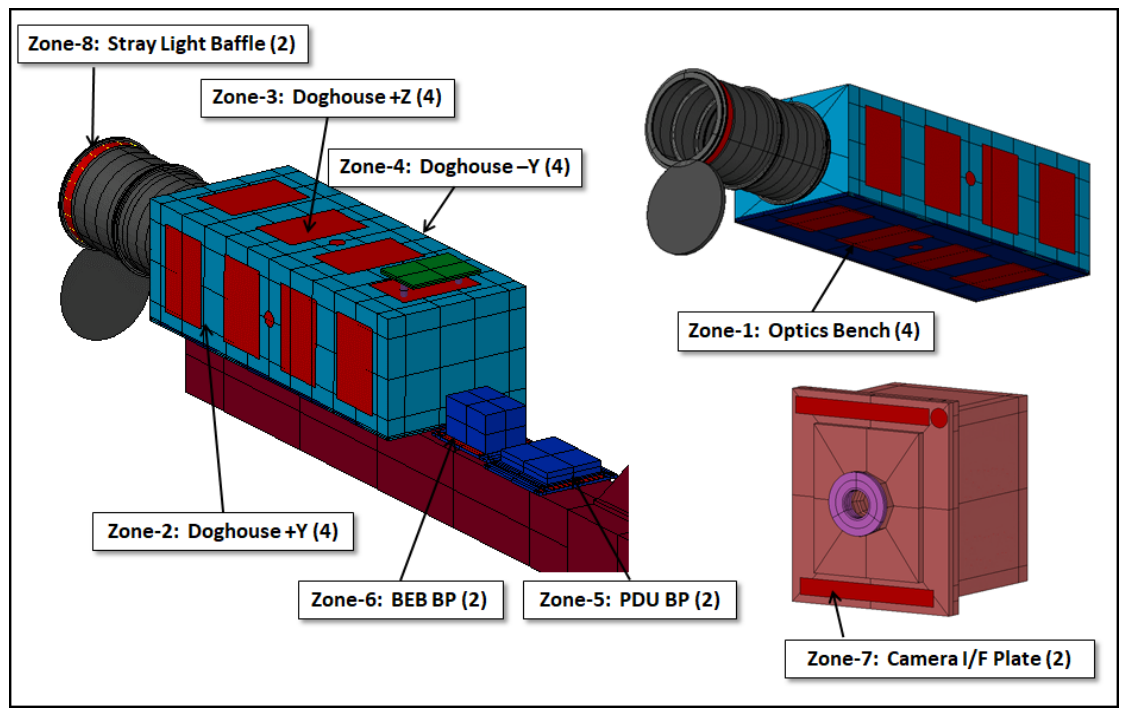

**Figure 8**. Thermal zones and the location of heaters (shown in red). The number of heaters in each zone is indicated in parentheses. The red dots denote the locations of thermal sensors. The camera housing with the interface plate is shown as an inset (it is located inside the doghouse at the aft end directly below the radiator shown in green).

**5 Flight Software**

The flight software (FSW) running on BEB is dedicated for subsystem control, system resource management, observation management, data handling, and automation. The FSW is based on the core Flight System (cFS), which is a platform- and project-independent, reusable software framework along with a set of reusable software applications. The FSW development took into consideration of the limited communication speed and the need for automatic operation without ground station contact. The FSW development preceded the subsystem readiness using hardware and software simulators at different stages.

The FSW architecture has the following three layers: the operating system (OS), cFS Middleware, and applications. Linux/Ubuntu OS manages the mission hardware and software resources. The Middleware consists of (i) the OS abstraction layer that isolates the FSW from the OS, (ii) the platform support package (PSP) that adapts FSW to the specific hardware and OS platform, and (iii) the core Flight Executive (cFE), which is a set of mission-independent, reusable, core flight software services and operating environment. cFE provides common services such as time, executive, event, and table services along with the software bus to FSW. The applications are of two types: those reusable from cFS and the ones developed specifically for the BITSE mission. The FSW block diagram in Fig. 9 shows how various applications access BITSE hardware and also WASP and SIP (CSBF).

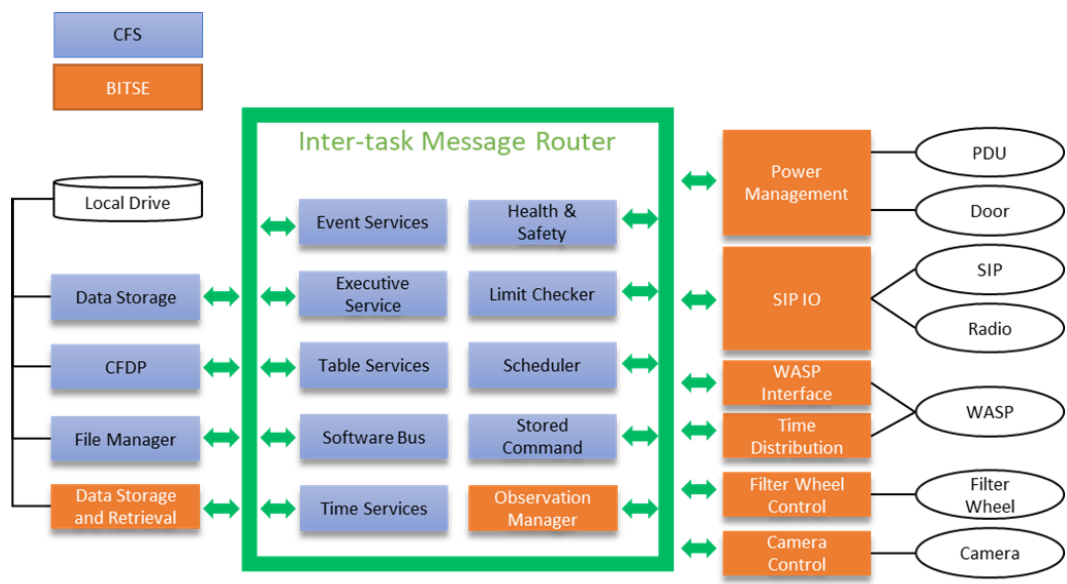

**Figure 9.** The BITSE FSW block diagram showing core Flight System (cFS) and BITSE applications and their connectivity to BITSE hardware. BITSE flight software uses the Consultative Committee for Space Data Systems (CCSDS) File Delivery Protocol (CFDP). The five core Flight Executive (cFE) services are shown in the left column inside the inter-task manager router. The remaining cFS blocks represent the reusable applications. The eight BITSE blocks (orange) represent applications developed specifically for the BITSE mission.

**6. Ground Support Equipment**

The ground support equipment (GSE) consists of a science computer, the WFF/WASP computer, and the Columbia Scientific Balloon Facility (CSBF) system, as well as the line-of-sight (LOS) command transmitter, the LOS Mini-SIP receiver, the LOS 1 Mbps receiver, and the LOS video receiver for WASP (see Fig. 10). The ground computer sends commands to the flight computer via the CSBF command transmitter. Upon execution of the commands, the data are sent down to be accessed by the ground science computer. Science data are downloaded using an 8 Mbps Ethernet-Via-Telemetry (EVTM) downlink.

The science ground system developed for BITSE leveraged heritage and open-source solutions used by previous GSFC flight missions. The ground system consists of three components: the COSMOS (Command and Control of Space Systems) command and telemetry system, the front-end data system, and the interface controller. The COSMOS ground system is an open-source application developed by Ball Aerospace which provides commanding, scripting, telemetry display, telemetry monitoring, and logging for ground and flight operations. The front-end data system provides data storage, retrieval, and real-time distribution and is based on a heritage application developed at GSFC for flight missions. The interface controller is an application that both simulates and interfaces to the CSBF facility ground station and is used both in ground testing as well as flight operations to bridge the user to physical interfaces to the balloon observatory.

The data analysis software on the image analysis computer provides many utilities to analyze and evaluate the images. It can display FITS image header information such as date, time, filter, exposure time, and so on. It also displays the image details such as size, pixel resolution, and cursor position. Intensity profile of the image can be displayed in the vertical and horizontal directions and both log and linear scales can be selected in displaying the profiles. The image can be displayed full screen and zoomed in and out. Facilities such as grid, scale-mark, concentric circles, and coordinates are used to assess the images. It is possible to select a region of interest (ROI) for a detailed look.

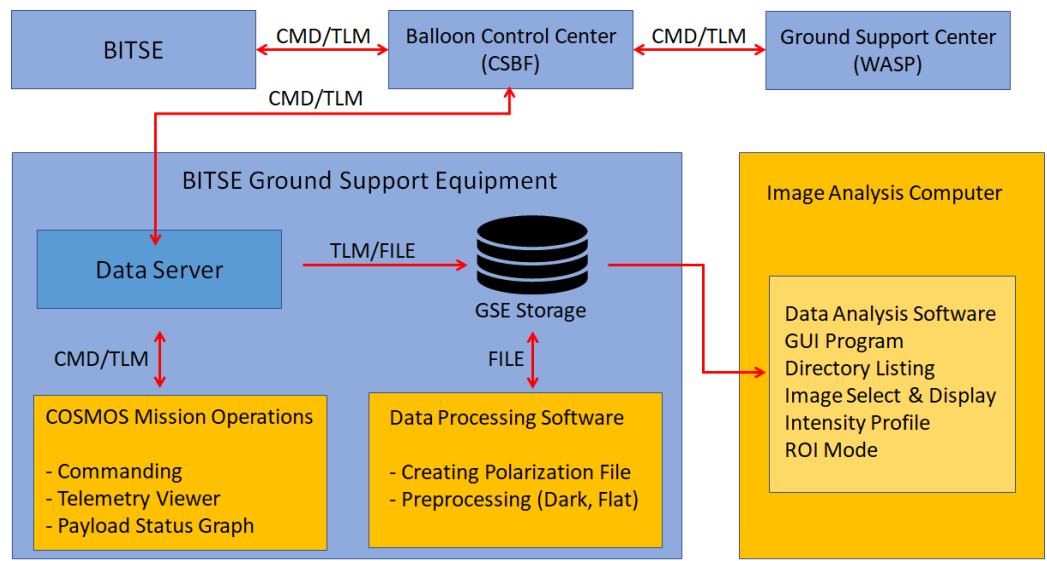

**Figure 10.** The block diagram of the Science Ground Support Equipment (GSE). The image analysis software helps monitor observations by downloading images and analyzing them, e.g., to check if the exposure time is appropriate. Raw images as well as the polarization images can be displayed and analyzed using the software tools.

## 7. Power Distribution Unit

Figure 11 shows the BITSE PDU that supplies power to BEB, polarization camera, FWA, aperture door, LOS transmitter, and various heaters for BITSE thermal control (see Section 4). The main source of power is the CSBF battery pack that supplies a nominal voltage of 30.8 V (range: 29.15 – 33 V) to the PDU via front input connector. PDU is designed to survive any input in the range of 0 to +40V DC and any unannounced removal of input power. PDU provides telemetry on the ±12V DC-DC converter thermistor, output voltage, and current monitors. The PDU is powered on during launch, thus there is no in-rush requirement. The PDU is mounted on a base plate, which in turn, is mounted on the telescope beam. It is thermally isolated from the telescope beams using struts. An input electromagnetic interference (EMI) filter and a common mode choke (CMC) are used at the DC-DC converter (±12V, 40W housekeeping power) inside the PDU. The operating temperature of the PDU is in the range -10ºC to +40ºC, while the survival temperature is in the range -20ºC to +50ºC.

The PDU provides unswitched and switched services. Unswitched power is provided to the polarization camera, the aperture door (one-time opening), the filter wheel assembly, thermal elements (heaters, thermistors, thermostat controls), and the CSBF Science Stack. Details can be seen in the block diagram (Fig. 11). Switched power is distributed to BEB and the LOS transmitter via latch relays. The CSBF Science Stack sends two 100-ms pulses to turn ON/OFF power source to the BEB and LOS transmitter.

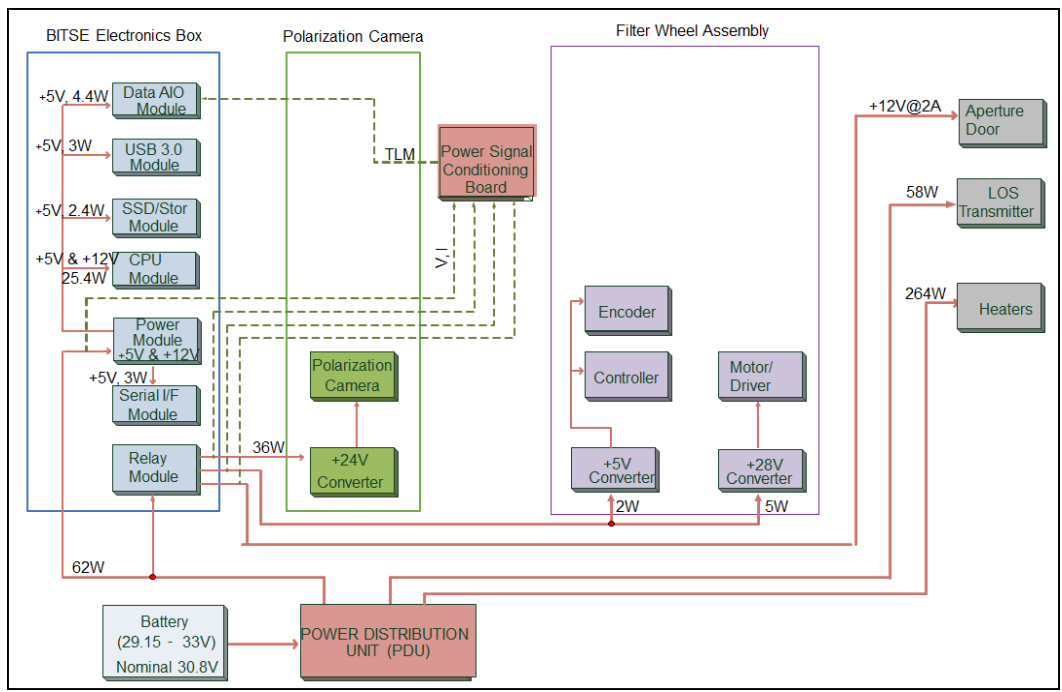

**Figure 11.** The block diagram of the power distribution unit (PDU), which takes input from the CSBF battery pack and distributes to all the subsystems of the coronagraph that use electric power. The estimated peak power for the mission is 465 Watts, and the estimated total battery consumption is 256 Amp-hours.

## 8. BITSE Integration and Launch

The filter wheel, camera, and BEB manufactured and tested at KASI, then received at GSFC and integrated with the optical and mechanical components in a clean tent. After alignment, the coronagraph was shipped to WFF where it was integrated with the gondola that provides the pointing mechanism needed for BITSE observations. Environmental tests have been performed at various levels from the subsystems to end-to-end. The Wallops Arc Second Pointing (WASP) system (http://sites.wff.nasa.gov/balloons/technology_wasp_details.html) provides < 1 arc sec pointing accuracy and sub arcsecond pointing stability, much more than required by BITSE. The pointing is achieved by a rectangular, hollow telescope beam mounted on a gimbal system controlled by two opposing gimbal hubs per axis (pitch and yaw). The telescope beam is sturdy enough to provide structural stability during flight and absorb any significant forces during landing. The gondola and the WASP system have been successfully used in previous test and science flights. The WASP system, the WASP avionics deck, the CSBF battery packs, and the science stack are accommodated on the gondola. The WASP avionics consists of a ≈21" × 17" (53 cm × 43 cm) aluminum deck, the main flight computer, enclosures for resolver interface, power relays, motor driver interface, GPS receiver, H-Bridge circuits (circuits graphically resembling the letter H enabling the motor to rotate in the forward and reverse directions), and housekeeping.

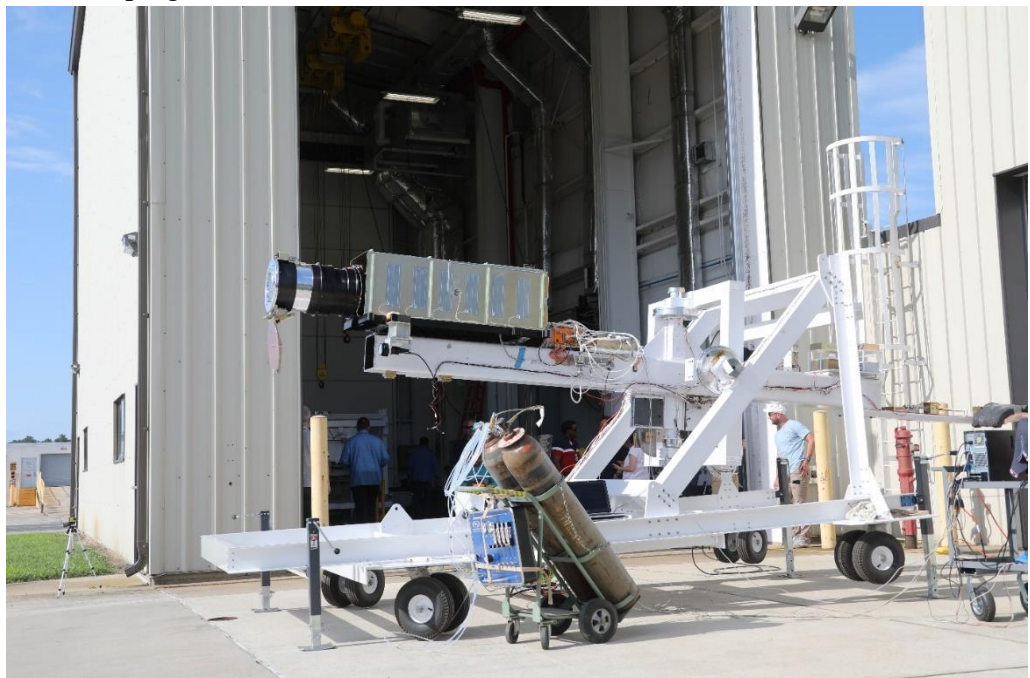

**Figure 12**. The coronagraph integrated to the gondola and WASP system (the telescope beam, gimbals, Sun sensor, and the associated computer) at WFF. The doghouse is placed over the coronagraph. The gondola was rolled outside the hangar for a Sun-pointing test. The nitrogen gas tanks in the foreground are for pushing the purge gas through the coronagraph system to keep the optics elements clean. The gondola is attached to tie ropes that connect to the rotator and then to the balloon.

The integration involved the following major steps: (i) attach the coronagraph on the telescope beam of the gondola such that the front baffle extends to the front of the beam, (ii) place and secure the BEB and PDU boxes behind the coronagraph on the telescope beam with base plates, (iii) complete the harnessing, (iv) fasten the WFF Sun sensor to the side of the optical bench. Figure 12 shows the coronagraph integrated to the gondola and the WASP system.

The gondola with all CSBF hardware had a total mass of ≈3200 lb (≈1450 kg). Adding the science payload mass, secondary payloads, steel ballast (≈900 lb or ≈408 kg), and the parachute the total mass became ≈5500 lb (≈2500 kg). Given the fact that the atmospheric scattering of sunlight is severe at the blue end, we chose the balloon that attains the highest possible altitude. Therefore, we chose the 39 million cubic feet (1.1 million cubic meter) balloon that attains a float altitude of ≈130,000 feet (≈40 km). The mission was ready for launch in early August 2019, but the launch happened only on September 18, 2019 when the weather was appropriate. The helium-filled balloon lifted off at 14:50 UT. The balloon attained the float altitude at 17:40 UT, after ascending for almost 3 hours. The aperture door was opened at 17:50 UT. After testing all functionalities, the observing started at 18:45 UT. Detailed steps from the beginning to the end of the BITSE mission is illustrated in Fig. 13.

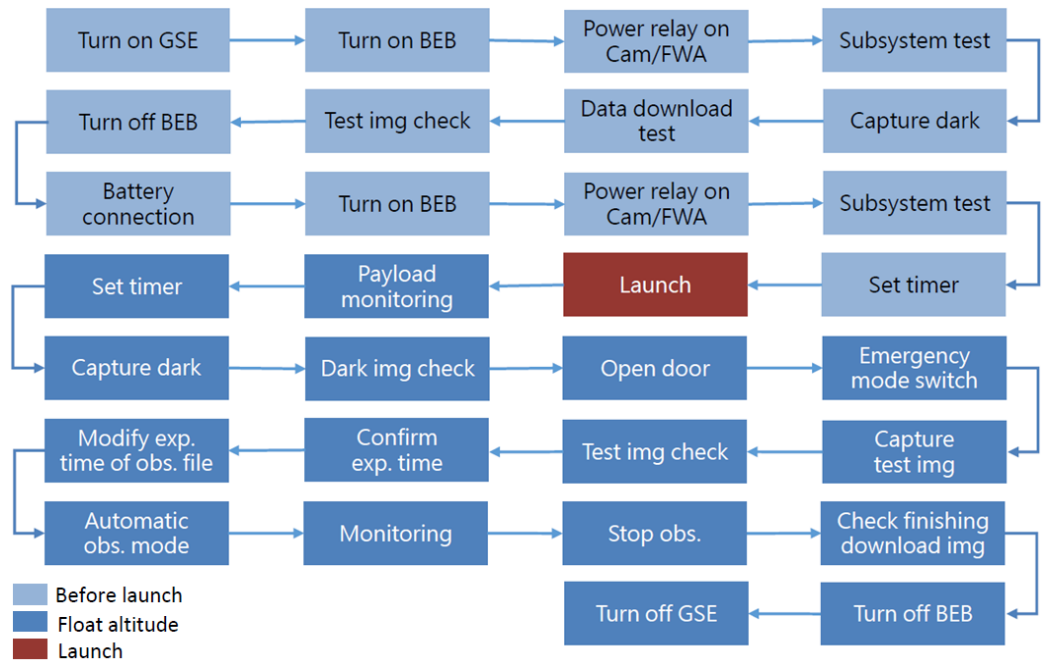

**Figure 13**. Activities before, during and after launch.

After an initial eastward excursion, the balloon moved in the northwest direction from Fort Sumner and passed between Santa Fe and Albuquerque. Towards the end of the observation, the WASP system failed, and the observations were stopped at 22:20 UT. The balloon was cut down and landed in the Navajo Nation near Farmington, New Mexico at 00:35 UT on the next day. The payload was recovered and trucked back to Fort Sumner two days later. All data were retrieved from the BEB. Table 4 lists the number of images obtained. Seventy frames were obtained in each narrow-band filter and two frames in the broadband filter per observing set. There were 61 sets in Filter 1 (F1) and F2 and only 60 sets in F3, F4 and F5, yielding 17,060 coronal images over an observing period of ≈4 hours. The cadence of the sets is ≈3.5 min. For dark images, 40 frames were obtained in each filter per set. Two dark sets were obtained at flight altitude, yielding 400 dark images.

Table 4 The number (N) of solar and dark images obtained by BITSE in each filter

| Filter N | Central λ (nm) | Sun frames | | Dark frames | |
|---|---|---|---|---|---|
| | | Exp. (s) | N Frames | Base Exp. (s) | N Frames[a] |
| 1 | 393.5 | 0.8 | 70×61 | 0.4 | 10×4×2 |
| 2 | 405.0 | 0.6 | 70×61 | 0.3 | 10×4×2 |
| 3 | 398.7 | 0.6 | 70×60 | 0.3 | 10×4×2 |
| 4 | 423.4 | 0.6 | 70×60 | 0.3 | 10×4×2 |
| 5 | 405.0 | 0.3 | 2×60 | 0.4 | 10×4×2 |

[a]10 frames each obtained for 4 different exposure times (base exp., 2×, 3×, and 5×base exp.) for 2 sets.

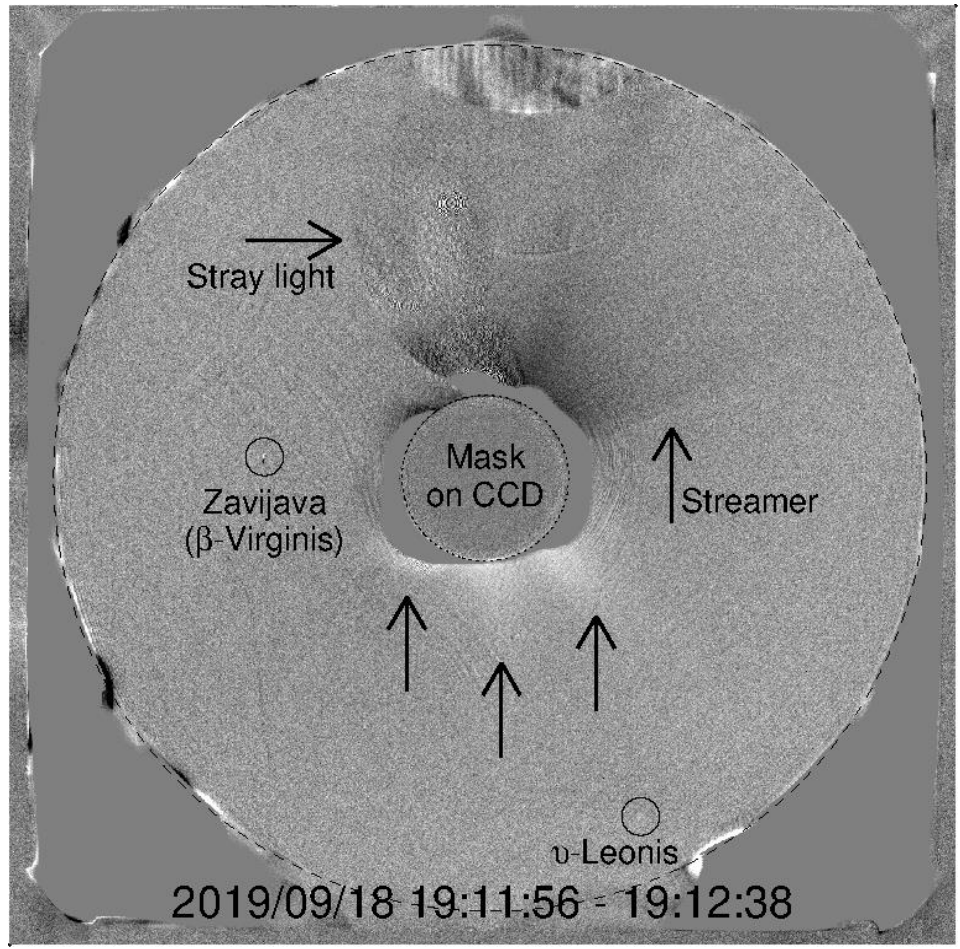

**Figure 14**. BITSE total brightness image in filter 4 at 19:11:56 UT on September 18. The signal from the coronal streamer (green arrow) is quite weak. The stray light is strong in the north and south (pointed by red arrows). There is also a region of saturation surrounding the image of the occulter on the polarization mask. Two stars Zavijava and υ-Leonis are shown circled. The star positions help us determine the Sun center in the image and the final pixel resolution. What is displayed is the ratio of the total brightness in set 7 to that in set 6. In each set, there are seventy images, which are added together.

## 9. Data Analysis and Preliminary Results

A 2048 × 2048 raw BITSE image is a composite of the four polarization positions of the MPA. Dark images, obtained with the same exposure time as the coronal images in a given filter, are summed together, and subtracted from the raw image. A sky image obtained through the coronagraph on August 14, 2019 is used for flat fielding and vignetting correction of the raw image. The individual polarization images are extracted from the raw image, giving four 1024 × 1024 images corresponding to the polarization positions $I_0$, $I_{45}$, $I_{90}$, and $I_{135}$.

Figure 14 shows a total brightness ratio image that combines all four polarization images. In creating the ratio image, the image of data set 7 (70 images taken during 19:11:56 – 19:12:38 UT co-added) has been divided by the image of data set 6 (70 images taken during 19:08:26 – 19:09:08). The image shows several features within the BITSE FOV. The streamer on the west side is the only obvious coronal signal. Also seen are two stars Zavijava (β Virginis, RA: $11^h\ 50^m\ 41.71824^s$ and Dec: +1° 45′ 52.9910″ at J2000.0) to the east of the Sun and upsilon Leonis (υ Leo, RA: $11^h\ 36^m\ 56.92983^s$ and Dec: +00° 49′ 25.8758″ at J2000.0) to the south. The relative positions of the two stars and the Sun are used to determine the Sun center and the pixel resolution. The Sun center is offset from the WASP pointing by ≈0.63 Rs. The pixel size in the final 1024 × 1024 images is 27.4″ compare to the designed 29.6″. The change in the pixel size seems to be due to a slight change in the effective focal length that occurred during the assembly of the detector.

Zavijava appears as a black-white pair due to the rotation of the BITSE FOV because of the alt-azimuth tracking system. The dark portion is the star's location at the time of data set 6, while the bright portion is the location corresponding to the time of set 7. The streamer on the west side shows a similar black-white structure in accordance with the clockwise rotation of the FOV.

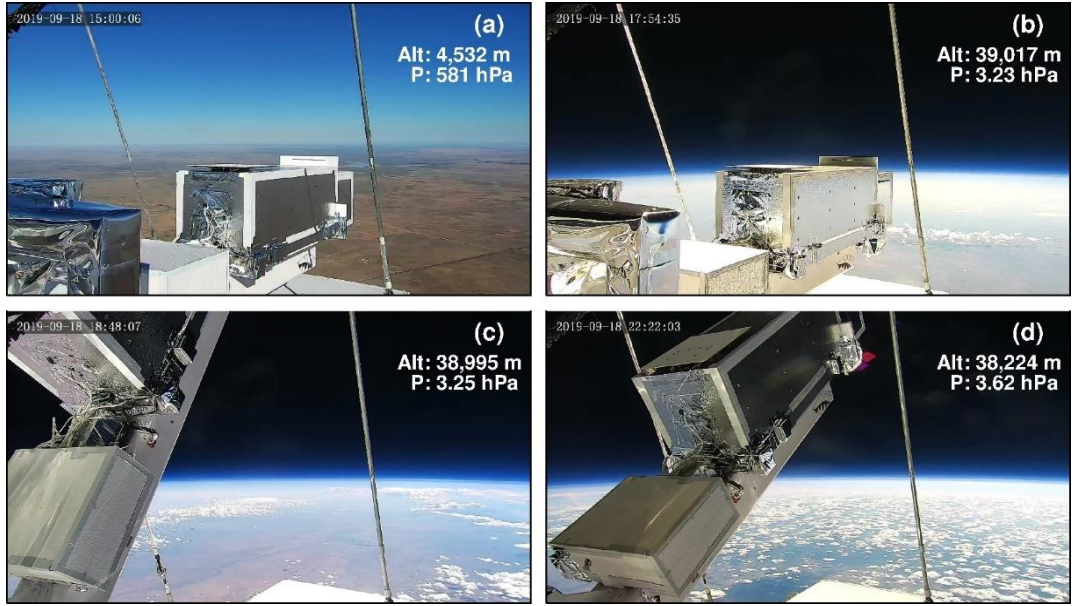

**Figure 15.** Snapshots of the coronagraph and its surroundings made by the onboard video camera. (a) near the ground level as the balloon lifts, (b) at 17:54 UT when the gondola reaches the float altitude, (c) at 18:48 UT when the coronagraph is pointed to the Sun, and (d) at 22:11 UT close to the end of the flight. The altitude (Alt, in m) of the gondola and the local atmospheric pressure (P, in hPa) are marked

The FOV also contains a ghost feature and scattered light in the instrument. A major contribution to the background is the sky brightness, which is not completely eliminated at the float altitude. Figure 15 shows snapshots of the coronagraph and its surroundings taken by the on-board video camera. Even though the sky looks dark at float altitude, there is still significant sky brightness level. The diffraction stray light also equally contributes to the background. The main problem in coronagraphic images is subtracting the background. Since we need to measure the polarization brightness to isolate the K-corona, we make use of the field rotation to isolate the signal from the streamer.

**9.1 Background Subtraction and Polarized Brightness**

The 1024 ×1024 images are rebinned to 256 × 256 for further processing. In this process, hot pixels are removed in each of the new macropixels. We subtract the background from polarized brightness (pB) image as follows. We track the Stokes variables $S_1 = I_0 - I_{90}$ and $S_2 = I_{45} - I_{135}$ as a function of time in each pixel in the 256 × 256 images. The pixels that are crossed by the streamer (e.g., the pixel denoted by a diamond in Fig. 16a) show an enhancement during the streamer transit (due to FOV rotation). Before and after the streamer transit, the intensity in these pixels is close to the background, similar to the pixel denoted by a square symbol in Fig.16a. Repeating this procedure for every pixel in the image, we get the pB image as

$$pB = (S_1^2 + S_2^2)^{1/2}. \quad (1)$$

Figure 16b shows $I_0$, $I_{45}$, $I_{90}$, and $I_{135}$ at the diamond pixel in Fig. 16a. Because the sky brightness increased due to the decreasing balloon altitude, all intensities increased with time. It is difficult to see the streamer signals in Fig. 16b because the background (sky and the stray light) was not constant. However, the streamer signals are clearly seen in the $S_1$ signal (= $I_0 - I_{90}$) as shown in Fig. 16d as the change of the background $S_1$ signal was relatively small. The streamer signal is not seen in $S_2$ (= $I_{45} - I_{135}$) but this is expected because the signal is vertically polarized at the streamer PA = 270⁰ (PA is the position angle and the polarization angle is tangential).

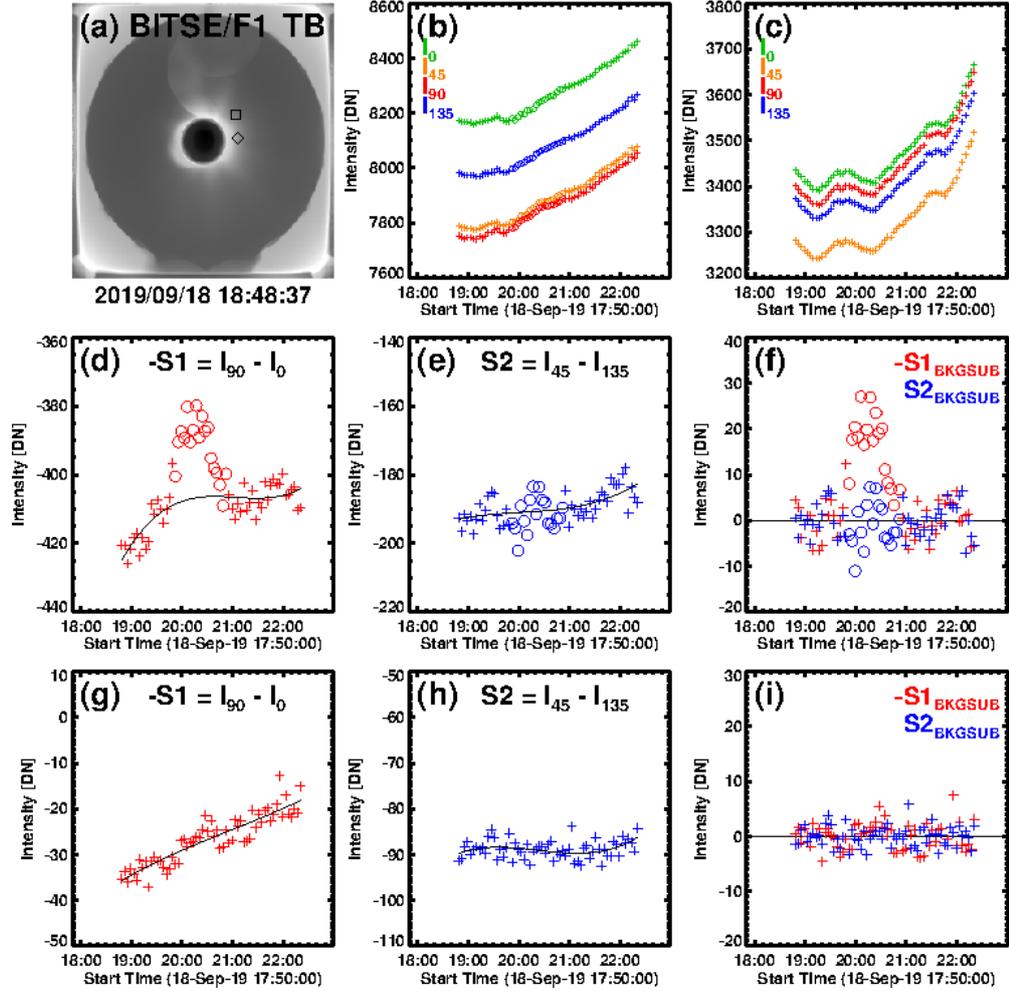

**Figure 16**. Background subtraction technique. (a) a total brightness image at 18:48:37 UT in filter 1 (F1) with two pixels marked with diamond and square symbols located on and outside the western streamer rotation range. The image corresponds to the first set of 70 images co-added and rebinned to a size of 256×256 pixels from the original size of 1024×1024 pixels. (b) F1 intensity in data number per second (DN/s) at the "diamond" pixel in the four polarization positions ($I_0$, $I_{45}$, $I_{90}$, and $I_{135}$). (c) F1 intensity at the "square" pixel in the four polarization positions ($I_0$, $I_{45}$, $I_{90}$, and $I_{135}$). (d) Stokes variable -$S_1$ = $I_{90} - I_0$ plotted as a function of time at the "diamond" pixel. The circles correspond to data points on streamer, while the "+" points correspond to the background (outside the streamer). A third order polynomial fit to the background points is superposed. Similarly, Stokes $S_2$ = $I_{45} - I_{135}$ is shown in (e) along with the third order

polynomial fit to the background data points. (f) $S_1$ and $S_2$ after subtracting the fitted background (denoted with subscripts BKGSUB). $S_1$ and $S_2$ change with time. When the fitted background from (d) and (e) are subtracted, the background values of $S_1$ and $S_2$ are at zero with small fluctuations. The procedure is repeated for the "square" pixel in (a) as shown in (g), (h), (i).

To estimate the background values of $S_1$ and $S_2$ signals during the streamer transit, we fit third-order polynomials to the off-streamer data (plus symbols) and subtract the fitted values from the $S_1$ and $S_2$ signals as shown in Fig. 16f. Figure 16c shows the variation of $I_0$, $I_{45}$, $I_{90}$, and $I_{135}$ at the square pixel in Fig. 16a. The western streamer did not transit this pixel during the observation. Figures 16g and 16h show the $S_1$ and $S_2$ signals of the square pixel along with the third-order polynomial fits. When the fitted values are subtracted from $S_1$ and $S_2$ we see the fluctuation of the background around zero (Fig. 16i).

We can use Fig. 16f to estimate the signal-to-noise ratio in the diamond pixel. The standard deviation of the off- streamer points (plus symbols in Fig. 16f) is 4.20 DN on S1 and 3.99 DN on S2. A similar noise level is expected in the on-streamer points (circles in Fig. 16f). Using the average S1 (-14.93 DN) and S2 (-1.18 DN) and the corresponding noise levels, we get the SNR in the pB signal as 2.6 according to Eq. (1). The SNR is low, but it corresponds to a snapshot image. We need to do temporal and spatial averaging of the images to improve the SNR. When we add 30 sets of images the SNR becomes 14.1. The SNR is further improved by rebinning the $256 \times 256$ image to a size of $32 \times 32$ pixels.

The background subtraction method outlined above is applied to all the pixels in the images taken at the four filters. After co-adding 30 sets of images, we obtain the pB images shown in Fig. 17. BITSE was launched at the end of Solar Cycle 24, so the corona has a simple structure with equatorial streamers above both limbs (marked 2 and 3 in Fig. 17). There was an additional thin streamer at the location marked 1, but this streamer is not discernible in the BITSE image. A closer look at streamer 2 indicates that there are striations in the lateral direction, which is likely to be due to insufficient flat-fielding. The offset between the occulter image (BITSE pointing center) and the optical disk was introduced in the beginning of the observations to make the features look symmetric. The offset is estimated as ≈0.63 Rs from the relative positions of the Sun, Zavijava and υ-Leo (see Fig. 14). This makes the western streamer observed well from 3 Rs onwards, while the eastern streamer correspondingly was observed only from ≈4 Rs. For these reasons, we primarily use the western streamer in determining the temperature and flow speed of electrons.

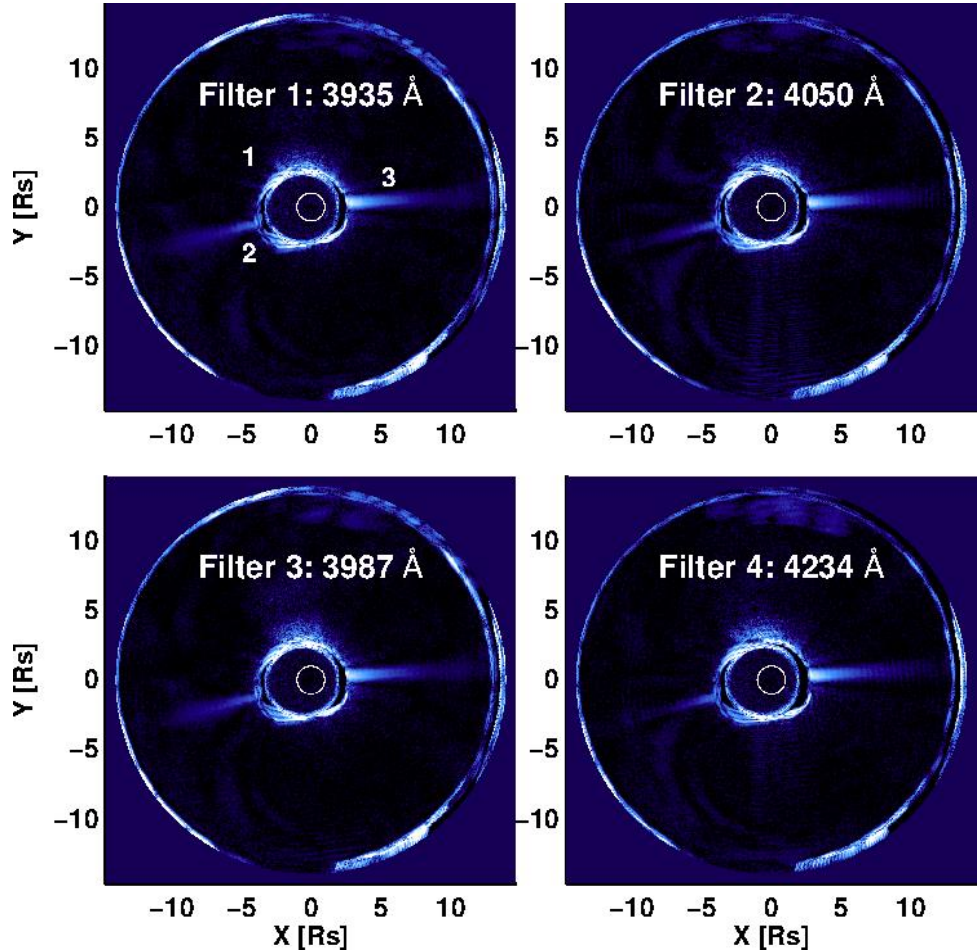

**Figure 17.** Polarized brightness (pB) images in the four narrowband filters. Thirty sets (sets 6 - 35) of 70 images each have been co-added to make these images. The innermost white circle represents the optical Sun. The next circle represents the focal plane occulter, which corresponds to the occulting mask placed on the micropolarizer array. The dark and bright features surrounding the occulting disk are due to saturation. The outermost circular boundary marks the edge of the outer FOV of BITSE COR.

**9.2 Comparison with LASCO images**

Figure 18 compares a background-subtracted BITSE/F1 pB image with a LASCO/C2 pB image, from which a monthly minimum background has been subtracted. The polarized brightness plotted against the PA at a heliocentric distance of 5 Rs, shows the streamers 2 at PA≈100º and 3 at PA ≈ 270º. Streamer 1 at PA ≈70º is discernible in the LASCO/C2 image, but not in the BITSE image. There is a hint of streamer 1 in the BITSE image, but almost at the background level. The LASCO/C2 image also shows polar plumes. The BITSE dynamic range is not high enough to show the polar plumes.

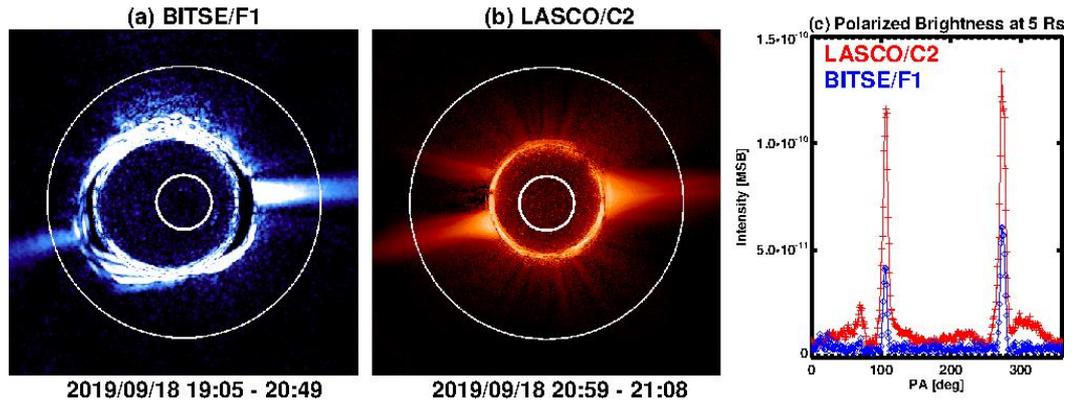

**Figure 18**. BITSE pB image in F1 (a) compared with the nearest LASCO/C2 pB image (b). A monthly minimum background of has been subtracted from the LASCO/C2 pB image that eliminates the F-coronal component. The time range over which the images were obtained is shown below the images. The BITSE/F1 (blue) and LASCO/C2 (red) pB as a function of position angle is plotted in (c) at a heliocentric distance of 5 Rs, indicated by the outer white circles in (a) and (b). Streamer 1 is clear in the LASCO image and profile, but barely noticeable in the BITSE profile. The BITSE profile is approximately calibrated making use of the fact that LASCO/C2 image was taken with an orange filter and our wavelengths are in the blue end of the photospheric spectrum and assuming that the K-coronal spectrum has similar wavelength variation.

Figure 19 shows the PA profile of streamers 2 and 3 in all four filters at 5 Rs compared with the LASCO/C2 profile. Also shown is the radial profile at PA = 275º. The LASCO/C2 pB intensity (in units of mean solar brightness, MSB) is high because it is through a passband at longer wavelengths (orange). Taking account of the relative locations of the four BITSE passbands and the LASCO/C2 passband (orange filter) on the photospheric spectra, we estimated the expected intensity at BITSE wavelengths. BITSE/F4 intensity is smaller by a factor of 0.937 than the LASCO/C2 intensity based on the photospheric spectrum. The intensity in the shortest BITSE wavelength is smaller by a factor of 0.520. The intensities in the other two wavelengths fall between these two limits. The signal becomes very weak beyond ≈7 Rs in the western streamer as indicated by the fluctuations in the BITSE radial profiles. We also determined LASCO/C2 pB profiles using the solarSoft ware pB_inverter routine. This profile is in agreement with the LASCO/C2 profile in Fig. 19c, even though the routine uses F-corona subtraction using the model by Koutchmy and Lamy (1985).

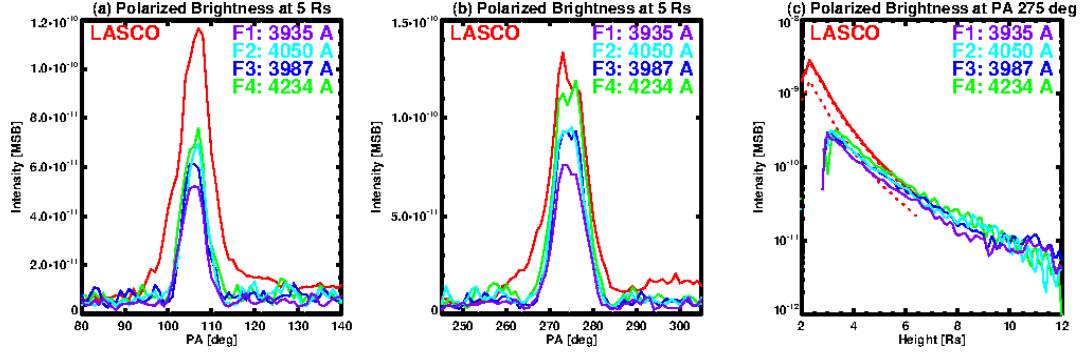

**Figure 19**. The position-angle profiles of streamer 2 (a) and 3 (b) at 5 Rs and the radial profile of streamer 3 (c). Intensities in all four filters are plotted. The red solid and dotted profiles in (c) are obtained by multiplying the LASCO/C2 profiles by 0.937 (corresponding to F4) and 0.520 (corresponding to F1), respectively based on the photospheric spectrum at LASCO and BITSE wavelengths. The inner FOV cut off is at 3 Rs for BITSE and 2.5 Rs in LASCO/C2, and hence the different profiles closer to the Sun. BITSE pB in DN/s are multiplied by $2 \times 10^{-11}$ MSB to match the LASCO pB.

## 9.3 Temperature and Speed of Coronal Electrons from the Spectral Fit

The technique to obtain coronal temperature (T) and speed (V) involves quantifying the shape change and red shift of the K-coronal spectrum (see e.g., Ichimoto et al., 1996). Since we have four data points on the spectrum, we can compare the observed spectrum against the theoretical spectrum to see to what T and V values give the best match between the two spectra. The computation of the theoretical K-corona spectrum requires T, V, and density (n) of electrons along the line-of-sight. Assuming T and V to be constant along the line of sight, we model the density distribution of the streamer using a density multiplier A given by

$$A = P_0 \exp\left[-\frac{1}{2}\left(\frac{\varphi - P_1}{P_2}\right)^2\right] \quad (2)$$

where $P_0$ is the peak density enhancement of the streamer above non-streamer region, $P_1$ is the central streamer location (longitude $\varphi = P_1$), and $P_2$ is the streamer width in longitude. For a spherically symmetric corona (no streamer), $P_0 = 0$ and $A = 1$. We computed $P_0$ as follows. We plotted the pB as a function of position angle at a heliocentric distance of 2.5 Rs (not shown). We determined the background level outside the streamers as $8 \times 10^{-11}$ MSB. The western streamer had a peak value of $\approx 1.6 \times 10^{-9}$ MSB, yielding $P_0 \approx 20$. We assume the same enhancement at different distances. The eastern streamer has a slightly higher enhancement factor ($\approx 25$), while the small streamer 1 has an enhancement factor of $\approx 5$. $P_1$ and $P_2$ are obtained from the time profiles of the polarized brightness in the LASCO/C2 synoptic maps (see Fig. 20). By fitting a Gaussian to these profiles, we get $P_1$ and $P_2$ as shown on the plots. The western streamer had a larger longitudinal extent. At the time of BITSE observation, the western streamer was on the west limb (only 0.26° below the sky plane). The eastern streamer was slightly behind the limb ($\approx 4.7°$), again not too far from the sky plane.

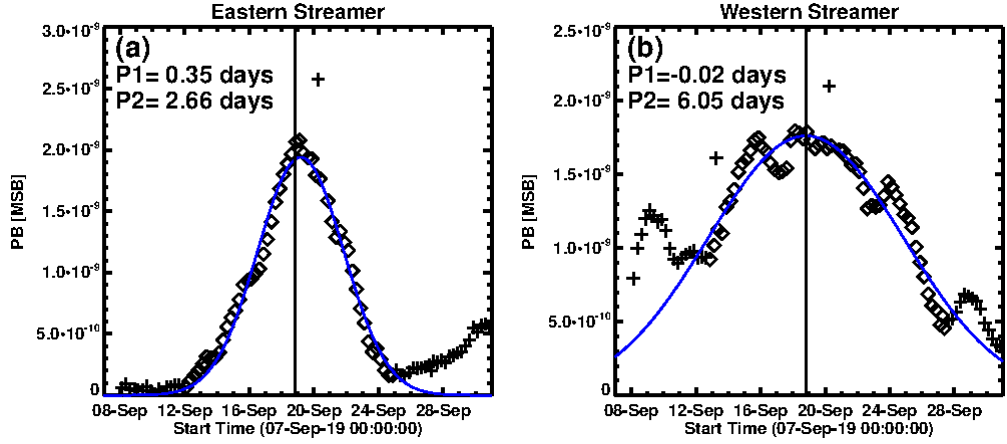

**Figure 20**. Polarized brightness (PB) vs. time plots of the eastern (left) and western (right) streamers in the height range 2.5 – 2.7 Rs from LASCO/C2 observations. The blue curve is the Gaussian fit to the measured pB values. P1 is expressed in days with respect to the position when the streamer was in the sky plane. For the eastern (western) streamer, the positive (negative) sign indicates that the streamer position is in the backside. P1 values of 0.35 days for the eastern streamer and -0.02 days for the western one correspond to ~4.7º and ~0.26º behind the limb on 2019 September 18 (BITSE flight).

The expected K-corona spectrum can be determined as described in Cram (1976) with the modification to include speed (Reginald and Davila, 2000). We use the input photospheric spectrum obtained on the day of BITSE observation from the Solar Irradiance and Climate Explorer (SORCE, Anderson and Cahalan, 2005) data available at http://lasp.colorado.edu/lisird/data/sorce_ssi_l3/. The spectrum is corrected for center-to-limb variation using known limb-darkening functions. There are several limb darkening models (Cram, 1976; Neckel and Labs, 1994), but we use Allen (1973) because different models do not seem to have significant effect on the K-corona spectrum. The coronal density distribution is taken to follow the Baumbach-Allen density model (Allen, 1947) multiplied by the factor A from Eq. (2). The Thomson-scattered radiation in the radial ($I_r$) and tangential ($I_t$) direction are then computed at each wavelength integrated along the line of sight. These are then combined to get the polarization and total brightness as ($I_t - I_r$) and ($I_t + I_r$), respectively. The resulting pB values at each wavelength is convolved with the CCD quantum efficiency and the filter profile. We also assume that the transmittance of the lenses and MPA are independent of the wavelengths over the narrow spectral range. Figure 21 shows the computed spectrum for a heliocentric distance of 5 Rs for various temperatures and assuming the flow speed to be zero. We used the QE and filter profiles provided by the manufacturers. The final expected spectrum we use has four data points corresponding to the four filter wavelengths.

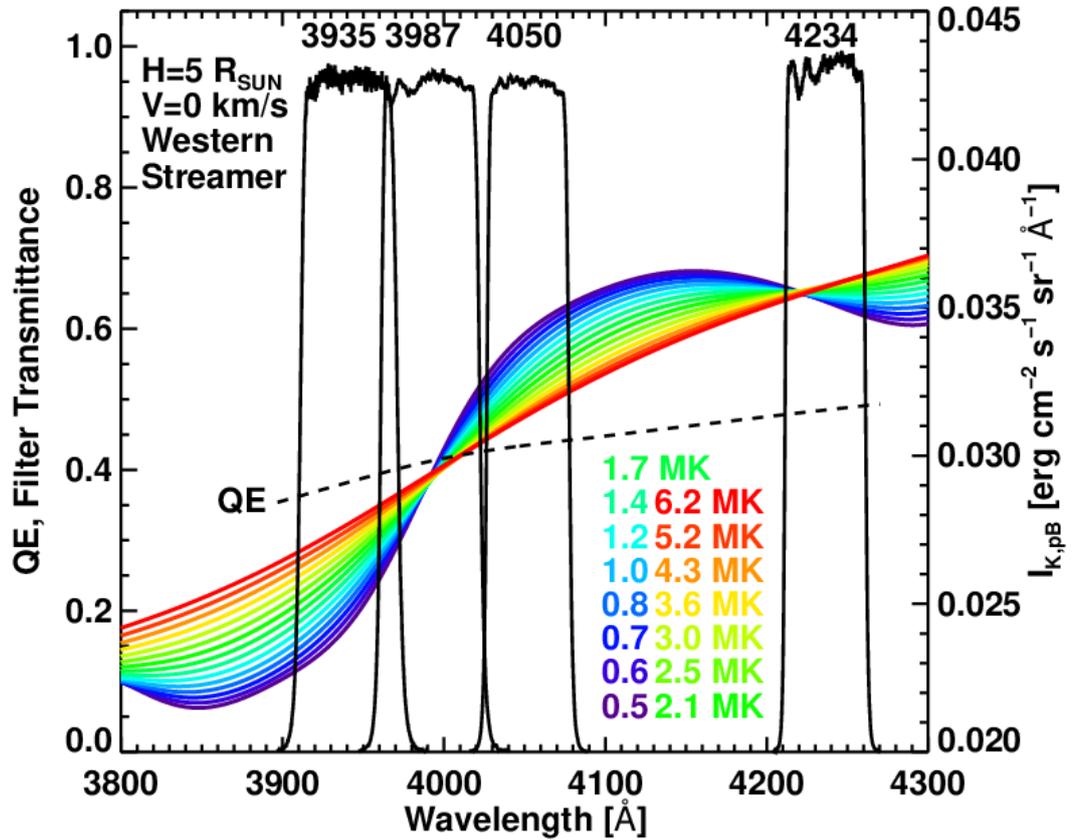

**Figure 21**. Expected spectrum of the K-corona in the western streamer at different temperatures as shown on the plot. The spectrum becomes smooth at higher temperatures because of the thermal motion of electrons. Also shown are the filter profiles (solid black curves) and the quantum efficiency (QE) of the CCD (dashed black line).

The only remaining information required for calculations is the atmospheric extinction because some light is lost due to scattering and absorption in the atmosphere above the float altitude. To correct for this, we first used the Simple Model of the Atmospheric Radiative Transfer of Sunshine (SMARTS) available online (https://www.nrel.gov/grid/solar-resource/smarts.html). We selected the SMARTS parameters so that the model estimated pressure matched with the air pressure measurements recorded during the BITSE flight. The atmospheric transmittance at BITSE float altitude computed using SMARTS is close to 1 at all wavelengths (3935 Å: 0.9955, 3987 Å: 0.9967, 4050 Å: 0.9969; and 4234 Å: 0.9971).

Another estimate of the atmospheric extinction is obtained making use of the observations of the star Zavijava within the BITSE FOV (see Fig. 14). We compared the intensity of Zavijava observed by BITSE at each filter wavelength with the spectrophotometric data (published by Kiehling, 1987; see also Glushneva et al., 1992) convolved with the filter profiles and the CCD quantum efficiency. The spectrophotometric data was listed at 5 nm resolution, so we linearly interpolated between the data points to match the higher resolution of the BITSE filter profiles. By comparing the observed and expected Zavijava spectrum, we get F2 and F3 wavelength transmittances with respect to that in the F4 wavelength as 1.0006 and 0.9837, respectively. At wavelengths < 400 nm, the observed Zavijava flux has large error (0.05 magnitude), so we assume the attenuation factor at the wavelength of F1 is the same as that of the nearby F3 wavelength. Therefore, to account for the spectral changes due to the atmospheric extinction, F1 and F3 intensities are divided by 0.9837. Spectral fitting using SMARTS and Zavijava extinction corrections yielded similar results, except that the former had a better minimum $\chi^2$. Therefore, we present the spectral fitting results using the SMARTS extinction.

Figure 22 compares the observed (black lines) and expected K-corona spectra (red lines) for different combinations of T and V at a macro pixel (32 × 32 original pixels) only in the western streamer located at a heliocentric distance of 4.2 Rs. The expected spectra shown in Fig. 21 are in physical units. To compare with the observed intensity in DN/s, we used the conversion factor as another free parameter. We used a least-squares method to find the best values of T and V for which the observed and expected spectra match as indicated by minimum $\chi^2$ ($\chi^2_{min}$) the goodness of the fit measure we used in fitting. T was varied between 0.2 and 6 MK, while V is varied between 0 and 800 km s$^{-1}$ to find the best T and V corresponding $\chi^2_{min}$. In Fig. 22a-c, we see that the observed and expected spectra do not match for any speed when T = 0.7 MK. In Fig. 22e, the two spectra match for T =1.0 MK and V =260 km s$^{-1}$. The two spectra deviate noticeably at lower (Fig. 22d) and higher (Fig. 22f) speeds. At higher T value (=1.4 MK in Fig. 22g-i) the observed and expected spectra do not agree for any speed. For T = 1.0 MK, we increased the speed from 0 km s$^{-1}$ to 800 km s$^{-1}$ and the full V range is consistent with this T, indicating a large uncertainty around V = 260 km s$^{-1}$. Fixing V = 260 km s$^{-1}$, we increased the temperature from 0.2 MK to 6 MK. The fit is reasonable for T between 0.7 and 1.4 MK, with the best value at 1.0 MK. Movies showing the spectral comparisons are provided in the supplement. The supplementary movie "bitse_specfit_findbest.gif" illustrates the spectral fitting to obtain best values of T and V in the pixel (20, 16) at a height H = 4.2 Rs. T is increased from 0.25 MK in steps of 0.05 MK and the speed is increased from 0 km s$^{-1}$ in steps of 10 km s$^{-1}$ to compare the computed spectrum (red lines) with the observed spectrum (black lines). The movie stops when $\chi^2$ attains its minimum value ($\chi^2_{min}$ = 0.009) for T = 0.97 MK and V = 260 km s$^{-1}$. Fixing V = 260 km s$^{-1}$, T is varied from 0.2 MK to 6.0 MK to get the uncertainty range in T corresponding to $\chi^2_{min}$ + 4 (see the supplementary movie "bitse_specfit_tchange.gif"). Fixing T = 0.97 MK, V is varied between 0 and 800 km s$^{-1}$ to estimate the uncertainty in V corresponding to $\chi^2_{min}$ + 4 (see the supplementary movie "bitse_specfit_vchange_0.97MK.gif").

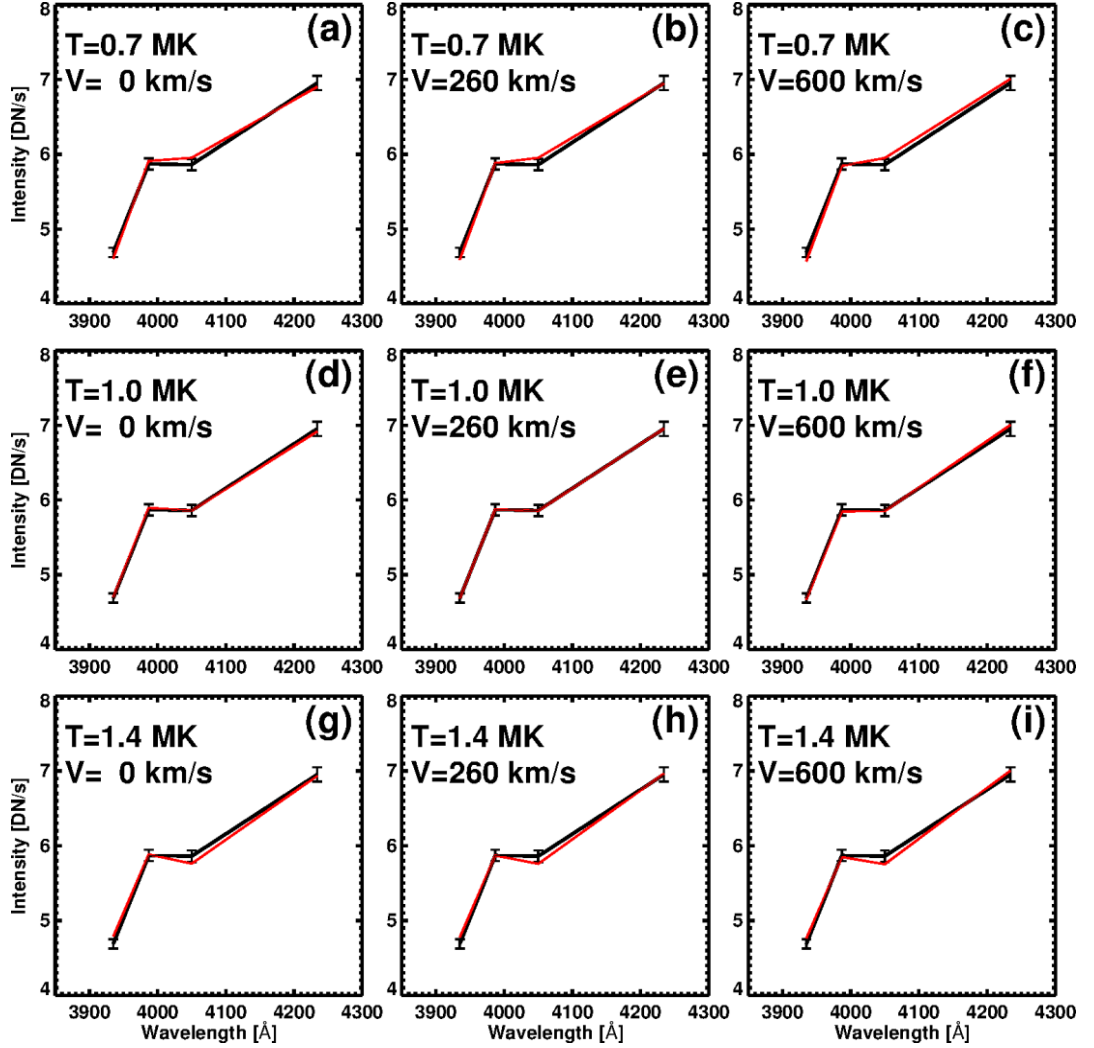

**Figure 22**. The observed K-corona spectrum (in black) compared with expected spectrum (in red) for various T and V combinations for a pixel on the streamer located at 4 Rs. T increases from the top to the bottom rows (0.7 MK, 1.0 MK, and 1.4 MK). V increases from the left to the right columns (0 km s$^{-1}$, 260 km s$^{-1}$, and 600 km s$^{-1}$). A connecting line is drawn between the data points, which are plotted at the central wavelength of the filter (Table 3). A complete range of T and V parameters is given in the supplementary movies (bitse_specfit_findbest.gif to determine best values of T and V, bitse_specfit_tchange.gif to determine the uncertainty range in T, and bitse_specfit_vchange_0.97MK.gif to determine the uncertainty range in V). The error bars in the observed spectrum are formal errors of rebinning the original 1024 × 1024 image to 32 × 32 size. Using the least squares method to minimize $\chi^2$, the best combination of T and V to represent the observation are found to be 1.0 MK and 260 km s$^{-1}$, respectively as seen in (e).

Figure 23 shows the temperature and speed values obtained using the method described in Fig. 22 at other pixels that have enough signal. Clearly, the signal is not strong enough on the eastern streamer to obtain reasonable T and V values. The western streamer has 5 pixels (19 – 23, 16). These pixels are marked in Fig. 23a by white boxes located approximately at heliocentric distances 3.3 to 6.9 Rs, in steps of 0.9 Rs. The innermost pixel (≈3.3 Rs) is bright in Fig. 23a, but it partly overlaps with the region of saturation. The signal is weak in the last two pixels located at 6.0 and 6.9 Rs. The spectral fitting method converges with $\chi^2_{min}$ = 0.009 and 5.022 in the 4.2 Rs (20, 16) and 5.1 Rs (21, 16) pixels. The corresponding best T and V values in these two pixels are (1.0 MK, 260 km s$^{-1}$) and (1.0 MK, 290 km s$^{-1}$). The confidence intervals are estimated by varying the parameters around the optimal values corresponding $\chi^2_{min}$ (see e.g., Bevington and Robinson, 2003). The 95% confidence intervals of T are determined from $\chi^2_{min}$ + 4 levels as (0.7 – 1.4 MK) and (0.7 – 1.3 MK) for these two pixels. In the last two pixels, the best T values are 1.1 ± 0.3 MK (6.0 Rs) and 0.9 ± 0.2 MK (6.9 Rs). If we fix the speed in the last two pixels to be 260 km s$^{-1}$ as in the 4.2 Rs pixel the T values remain the same. The speed range is very wide, between 0 and 800 km s$^{-1}$ in the last four pixels. We can say that the average temperature in the pixels between 4.2 and 6.9 Rs is 1.0 ± 0.3 MK. The flow speed in the height range is 260 km s$^{-1}$, but the uncertainty is large (0 – 800 km s$^{-1}$). Given the fact fast wind is not present in streamers, we can conclude that 260 km s$^{-1}$ is quite reasonable.

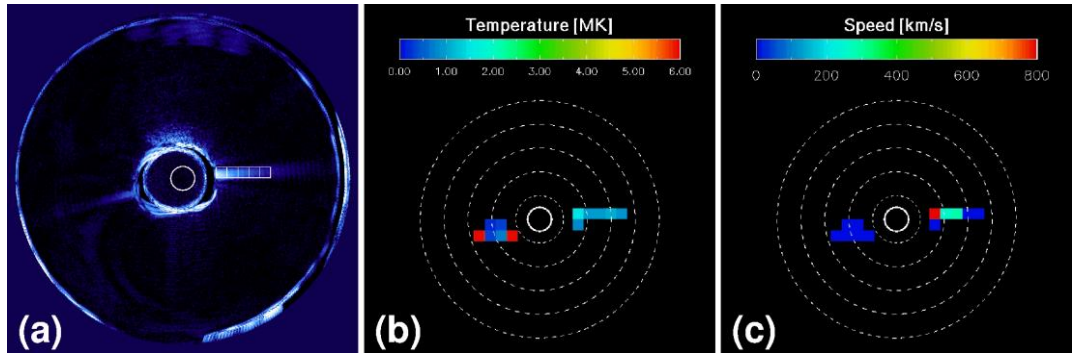

**Figure 23.** (a) Macro pixels of size 876.8×876.8 arc sec superposed on the western streamer in a pB image of size 256×256 pixels. Temperature and flow speed determined in these pixels are shown in the maps (b) and (c), respectively by the spectral fitting method outlined in Fig. 22. (b) and (c) have been rebinned to a size of 32×32 pixels. The marked pixels in (a) correspond to the row pixel 16 and column pixels 19-23. The spectral fitting was performed in each pixel of the western and eastern streamers to determine the best temperature and speed as displayed in (b) and (c). The red and blue pixels give extreme values, which are unphysical because the fitting did not converge. When we assume that these pixels have the speed of the adjacent pixel with good fit, the derived temperature is the same as in (b). The T values in pixels 19-23: 1.5, 1.0, 1.0, 1.1 and 0.9 MK. The best speeds are 260 and 290 km/s in pixels at 4.2 and 5.1 Rs, respectively. However, fixing V=260 km/s, the T values remain the same at $\chi^2_{min}$.

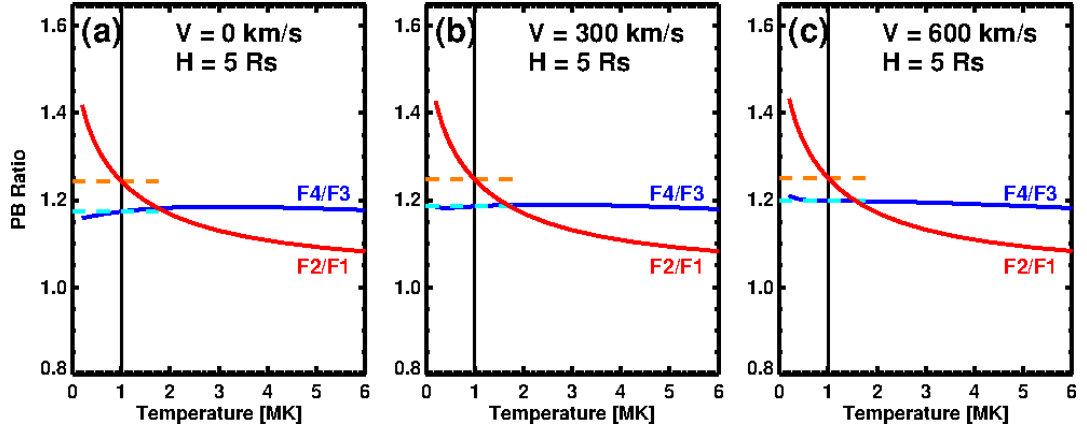

**Figure 24.** T (F2/F1) and V (F4/F3) ratios obtained from the expected spectrum are plotted as a function of T for three different speeds (0, 300, and 600 km/s). The T ratio corresponding to T = 1.0 MK is marked by the orange dashed lines, increases slightly with speed: 1.243 (a), 1.247 (b), and 1.250 (c). The V ratio corresponding to T = 1.0 MK is marked by the cyan dashed line also increases with speed: 1.174 (a), 1.186 (b), and 1.198 (c).

### 9.4 Ratio Maps

Figure 24 shows plots of the temperature (F2/F1) and speed (F4/F3) ratios as a function of T for three different speeds. The ratios were obtained using the expected pB computed as described in the beginning of section 9.3. Recall that the pixel (20,16) at ≈4 Rs on the western streamer has T = 1.0 MK obtained from the spectral fitting method. The temperature corresponds to a T ratio of 1.246 from the expected intensities at filters F2 and F1. The ratio falls smoothly as T increases. For example, as T increases from 1 to 2 MK, the T ratio falls from 1.246 to 1.169 for V=260 km s$^{-1}$. The drop in the T ratio by 7.7% per 1 MK, with a slightly faster rate near 1 MK than near 2 MK. The V ratio rises from 1.174 for 0 km s$^{-1}$ to 1.185 for 300 km s$^{-1}$ keeping T = 1.0 MK. The rise is by 0.40% for an increase of 100 km s$^{-1}$.

Figure 25 shows the filter ratio maps (F2/F1) and (F4/F3) along with the T and V maps derived from them. The F2/F1 map in Fig. 25a is much smoother than the F4/F3 map. This is especially true for the western streamer. The distribution of the ratios between 4 and 7 Rs in the western streamer gives an average T ratio of 1.25±0.025, which corresponds to T = 1.1 MK with a confidence interval of 0.9 MK to 1.3 MK. If we restrict the T ratio to the range shown in Fig. 24 (red line, 1.08 to 1.42), then the distribution has an average 1.249 ± 0.010. From Fig. 24, we see that this ratio corresponds to 1.0 ± 0.1 MK, in agreement with the spectral fitting method (see Fig. 23b). This value is an average over the radial distance 4 – 7 Rs. We can also convert the T ratio to temperature according to the curve in Fig. 24 in individual pixels. A histogram of the T values in the converted pixels has a slightly higher value: 1.1 ± 0.1 MK. In the eastern streamer there are some clusters of pixels having this temperature, but the signal-to-noise ratio is not good enough to obtain the temperature. The V ratio distribution is too broad to yield V values in the 256 × 256 image (Fig. 25d). Clearly, the maps are too noisy to infer the speeds. If we restrict the V the ratios to the range indicated by the blue lines in Fig. 24 and convert them to speeds, we get a distribution with an average value $297 \pm 66$ km s$^{-1}$.

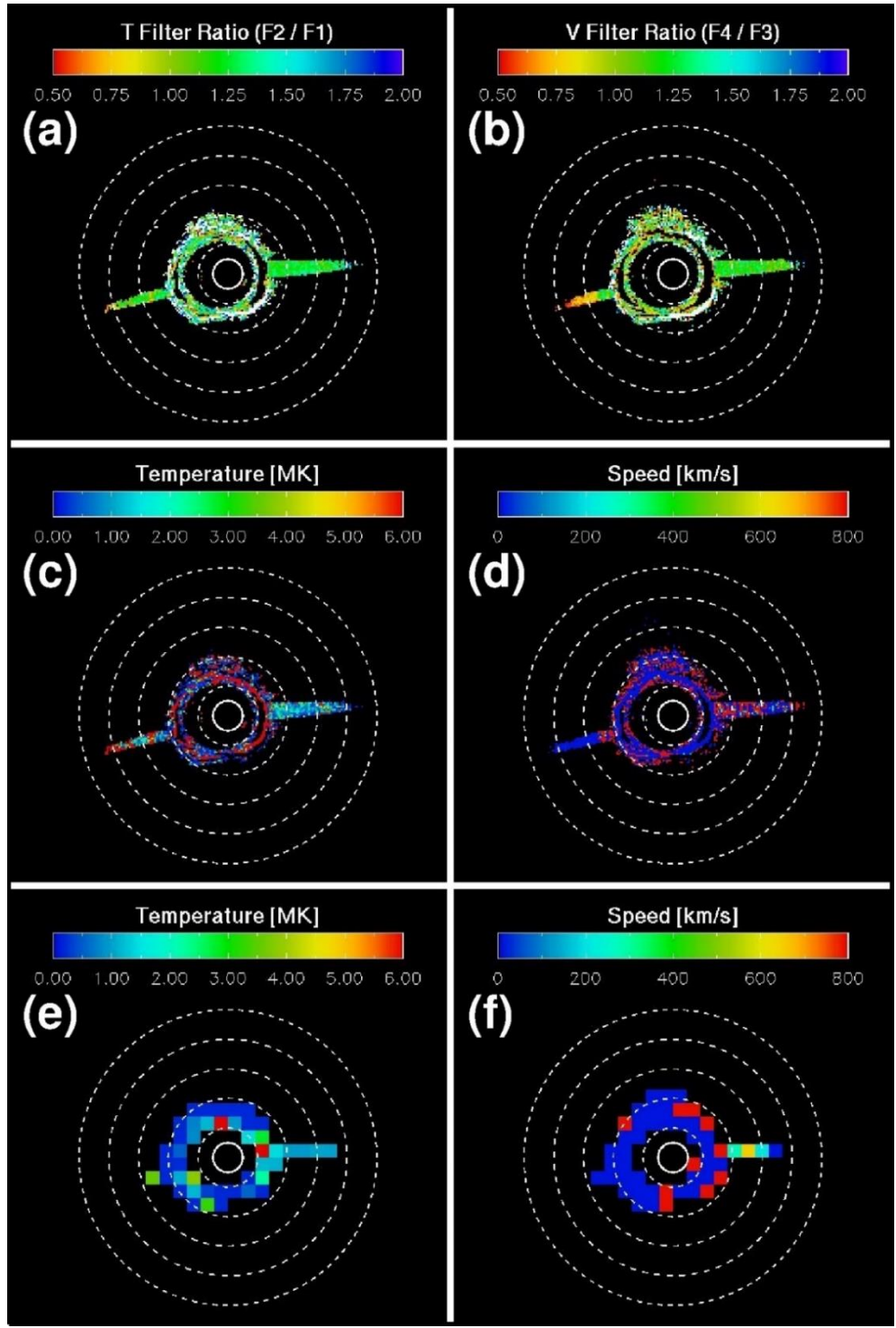

**Figure 25.** Filter ratio maps for temperature (F2/F1, a) and speed (F4/F3, b) along with the temperature (c,e) and speed (d,f) maps derived from the ratio maps. The maps (a,b) and (c,d) have each a size of 256 × 256 pixels, while the maps (e,f) have a size of 32 × 32 pixels. Regions outside the streamers where there is no signal have been masked out. In each image, the innermost solid white circle represents the optical Sun. The dashed circles are at heliocentric distances 2, 4, 6, 8, and 10 Rs. Note that the region < 4 Rs is in the saturation region.

Further spatial averaging (rebinning the pB image to a size of 32×32 pixels, obtaining the ratio maps, and converting them to T and V maps) shows that the temperature of the western streamer region to be at 1.0 MK in the four pixels at 4.2 -6.9 Rs (see Fig. 25e) in agreement with what was obtained from spectral fitting (see Fig. 23b). The V values in these pixels are 270 km s$^{-1}$, 650 km s$^{-1}$, 210 km s$^{-1}$ and 10 km s$^{-1}$ (see Fig. 25f). Clearly 5.1 Rs and 6.9 Rs pixels have extreme values, which are unphysical. In Fig. 25f, there are two pixels with speeds similar to those from spectral fitting: the pixels at 4.2 and 6.0 Rs. The pixel at 5.1 Rs had a speed of 290 km s$^{-1}$ from spectral fitting, but in the ratio image, this pixel has a much higher speed (650 km s$^{-1}$ bracketed by 270 km s$^{-1}$ on the left and 210 km s$^{-1}$ on the right). The speed of 650 km s$^{-1}$ corresponds to a V ratio of 1.200, which is only 2.6% higher than the ratio for zero speed (1.174). In the future work, we will improve the data analysis with refined calibrations and adding the remaining images.

**10. Discussion**

The objective of the BITSE mission was to fly a new single-stage coronagraph to image the corona in four narrowband filters in the blue end of the solar spectrum and to derive the physical parameters in the outer corona (3 – 8 Rs). The BITSE mission demonstrated that the coronagraph was able to obtain polarized brightness images using the polarization camera. The electron temperature and flow speed were derived from the pB images. The intensity of the K-corona dominates only at heliocentric distances < 2 Rs. Beyond 2 Rs, the F-corona is brighter than the K-corona. For example, at a distance of ≈3 Rs, the F-corona is brighter by a factor of ≈3. Fortunately, the F-corona is unpolarized up to ≈5 – 6 Rs corresponding to the inner part of the BITSE FOV. Thus, extracting the polarized brightness provides a true measure of the K-corona.

Due to the limited signal-to-noise ratio, we were able to determine the temperature and flow speed on the western streamer. We found that the electron temperature is ≈1.0 MK and the flow speed is ≈260 km s$^{-1}$ over the distance range of 4 – 7 Rs. We used spectral fitting and passband ratio imaging. Both methods give similar results for temperature, but the speeds have large uncertainties because the speed is extremely sensitive to the filter ratio. Dolei, Spadaro, and Ventura (2015) studied the properties of an equatorial streamer during Solar Cycle 23 minimum by combining visible light and ultraviolet observations. From the electron density derived from LASCO pB images, these authors estimated an electron temperatures in the range 0.6 – 0.9 MK at 4 Rs and 0.4 – 0.7 MK at 5 Rs. Their 4 Rs temperature overlaps with our range 0.7 – 1.4 MK; the 5 Rs temperatures barely overlap with the BITSE range (0.7 – 1.3 MK). On the other hand, electron flow speed is slightly larger than that derived from UVCS observations. It must be noted that the macropixels over which we determined the speed and temperature covers the whole streamer (see Fig. 23a). Therefore, we are unable to differentiate the speeds at the axis and edge of streamers.

The solar minimum streamer has a rather small spatial extent (≈10º in position angle). In the longitudinal direction, the western streamer lasts for ≈6 days. The same streamer structure was studied by Morgan and Cook (2020) two rotations after the BITSE observations. Their compilation of streamer densities at 5 Rs is in the range $(5 – 11)\times10^4$ cm$^{-3}$. We have not yet determined the electron density from the BITSE data yet, but we inverted the calibrated LASCO/C2 pB image (see Fig. 19) to get a density of ≈$7.6\times10^4$ cm$^{-3}$, which is within the above density range. After complete calibration, we plan to obtain the density from BITSE images. Assuming a constant mass flux and using the constraints on proton flux measured by Parker Solar Probe, these authors estimated an outflow in the streamers of 50 – 120 kms$^{-1}$ at 4Rs, and 90 – 250 kms$^{-1}$ at 8 Rs. Our speeds are slightly larger, but consistent with these estimates, given the large uncertainty in our speed measurements.

**11. Conclusion and Future work**

The BITSE mission has demonstrated that measuring the polarized brightness of the corona at several wavelengths and characterizing the shape of the corona is a viable technique to determine the physical properties of the corona in the solar wind acceleration region. Previous attempts to use this technique during solar eclipses sampled the corona much closer to the Sun. BITSE was able to obtain the temperature and speed in the range 4 – 7 Rs.

BITSE was launched on 18 September 2019, which corresponds to the minimum of Solar Cycle 24. The prominent features in the solar minimum corona are the equatorial streamers. The narrow streamer observed by BITSE is in good agreement with the geometric properties obtained by SOHO/LASCO in the overlapping physical domain. For the equatorial streamer on the west limb, we obtained a temperature of ≈1.0 ± 0.3 MK and a flow speed of ≈260 km s$^{-1}$. These parameters are in the general range of other estimates in the literature obtained indirectly. These results are preliminary, and we expect to refine them by (i) adding more images that need to be processed taking account of the slightly different BITSE altitudes, and (ii) completing post-launch calibrations.

One of the technologies we demonstrated is the use of the polarization camera used for the first time in space. Obtaining the polarized brightness using the polarization camera is advantageous over the traditional polarization mechanism that combines observations from three polarization positions. This is also important for future space missions because the polarization camera reduces one mechanism (polarization wheel). We were not able to test the tapered-cone external occulter because the occulter size manufactured was different from the design size. An older design value was used in the manufacture. We ended up using a rings attached to the occulter to bring it to the required size. We expect to use the tapered cone occulter in a follow-up BITSE flight. The follow-up flight will also overcome the shorter-than-planned duration of observation: the late start (due to wind conditions) and early end (due to failure of the WASP system).

BITSE is the first joint solar mission between NASA and KASI, a step to increase technology readiness level toward a next-generation coronagraph (Cho et al., 2017). Through BITSE, the NASA and KASI coronagraph teams demonstrated the technology successfully and are developing a a coronagraph for the Experiment COronal Diagnostic Experiment (CODEX) that will be deployed on the International Space Station 2023. CODEX will overcome the sky brightness and the short observing duration that resulted in insufficient signal-to-noise ratio beyond ≈7 Rs.

**Acknowledgements**
This work is supported by NASA's Technology Development Program and KASI's R and D program, Development of a Solar Coronagraph on the International Space Station. We acknowledge the coordinated effort by the engineering, WASP, and CSBF teams (optical: Lenward Seals, electrical: Hanson Nguyen, mechanical: Melvin Donahue, Jason Budinoff, Peter Patterson, ground system: Joseph-Paul Swinski, thermal: William Chang, WASP: Scott Heatwole, James Lanzi, David Stuchlik, CSBF: Robert Salter).

**Disclosure of Potential Conflicts of Interest:** The authors declare that there are no conflicts of interest

**References**

Abbo, L., Ofman, L., Antiochos, S. K., Hansteen, V. H., Harra, L., Ko, Y.-K., et al.:2016, Slow Solar Wind: Observations and Modeling. *Space. Sci. Rev.* 201, 55. DOI. ADS.

Allen, C. W.: 1947, Interpretation of Electron Densities from Corona Brightness. *Mon. Not. R. Astron. Soc.* **107**, 426.

Allen, C. W.: 1973, Astrophysical quantities. London: University of London, Athlone Press. P.170

Anderson, D. E., Cahalan, R. F.: 2005, The Solar Radiation and Climate Experiment (SORCE) Mission for the NASA Earth Observing System (EOS). *Solar Phys.* **230**, 3

Bevington, P. R., Robinson, D. K.: 2003, Data Reduction and Error Analysis for the Physical Sciences. Boston: McGraw-Hill, p. 194

Cho, K.-S., Bong, S.-C., Choi, S., Yang, H., Kim, J., Baek, J. -H. et al., 2017, Toward a Next Generation Solar Coronagraph: Development of a Compact Diagnostic Coronagraph for the ISS, *J. Korean Astr. Soc* **50**, 139

Cho, K.-S., Yang, H.-S., Lee, J.-O., Bong, S.-C., Kim, J., Choi, S. et al.: 2020, Toward a Next Generation Solar Coronagraph: Diagnostic Coronagraph Experiment, *J. Korean Astr. Soc.* , **53**, 87

Cram, L. E.: 1976, Determination of the temperature of the solar corona from the spectrum of the electron-scattering continuum. *Solar Phys.* **48**, 3. DOI. ADS.

Cranmer, S. R.: 2012, Self-Consistent Models of the Solar Wind. *Space Sci. Rev*. **172**, 145. DOI. ADS.

Cranmer, S. R., Kohl, J. L., Miralles, M. P., van Ballegooijen, A. A.: 2010, Extended Coronal Heating and Solar Wind Acceleration over the Solar Cycle, in SOHO-23: Understanding a Peculiar Solar Minimum ASP Conference Series Vol. 428, proceedings of a workshop held 21-25 September 2009 in Northeast Harbor, Maine, USA. Edited by Steven R. Cranmer, J. Todd Hoeksema, John L. Kohl. San Francisco: Astronomical Society of the Pacific, 2010, p.209

Cranmer, S. R., van Ballegooijen, A. A.: 2012, Proton, Electron, and Ion Heating in the Fast Solar Wind from Nonlinear Coupling between Alfvénic and Fast-mode Turbulence. *Astrophys. J.* **754**, 92. DOI. ADS.

Dolei, S., Spadaro, D., R. Ventura, R.: 2015, Visible light and ultraviolet observations of coronal structures: physical properties of an equatorial streamer and modelling of the F corona. *Astron. Astrophys.* **577**, A34. DOI. ADS.

Endeve, E., Holzer, T. E., Leer, E.: 2004, Helmet Streamers Gone Unstable: Two-Fluid Magnetohydrodynamic Models of the Solar Corona. *Astrophys. J.* **603**, 307. DOI. ADS.

Fineschi, S., Gardner, L. D., Kohl, J. L., Romoli, M., & Noci, G.: 1998, Grating stray light analysis and control in the UVCS/SOHO, *SPIE Proc.* **3443**, 67

Glushneva, I. N., Kharitonov, A. V., Kniazeva, L. N., Shenavrin, V. I.: 1992, Secondary spectrophotometric standards. *Astron. Astrophys.* **92**, 1

Gong, Q., Gopalswamy, N., Newmark, J.: 2019, Innovative compact coronagraph approach for Balloon-borne Investigation of Temperature and Speed of Electrons in the corona (BITSE), *SPIE Proc* **11116**, id. 111160F

Gopalswamy, N., Gong, Q.: 2018, A small satellite mission for solar coronagraphy. *SPIE Proc*. **10769**, id. 107690X. DOI. ADS.

Gopalswamy, N., Yashiro, S. Reginald, N., Thakur, N., Thompson, B. J., Gong, Q.: 2018, Physical Conditions in the Solar Corona Derived from the Total Solar Eclipse Observations obtained on 2017 August 21 Using a Polarization Camera, American Astronomical Society, *AAS Meeting* #**231**, id. 220.08

Grall, R. R., Coles, W. A., Klinglesmith, M. T., Breen, A. R., Williams, P. J. S., Markkanen, J., Esser, R.: 1996, Rapid acceleration of the polar solar wind. *Nature* **379**, 429

Ichimoto, K., Kumagai, K., Sano, I., Kobiki, T., Sakurai, T., & Munoz, A.: 1996, Measurement of the Coronal Electron Temperature at the Total Solar Eclipse on 1994 November 3. Publ. Astron. Soc. Jpn **48**, 545. DOI. ADS.

Ichimoto, K., Kumagai, K., Sano, I., Kobiki, T., Sakurai, T., Munoz, A.: 1997, Measurement of the coronal electron temperature at the total solar eclipse on 1994 November 3. NATO Advanced Research Workshop on Theoretical and Observational Problems Related to Solar Eclipses. *NATO ASI Series C*, Vol. 494. Dordrecht: Kluwer Academic Publishers, 1997, p. 31 – 34. DOI. ADS.

Kiehling, R.: 1987, Spectrophotometry of bright F-, G-, K- and M-type stars. I. Measurements of 60 southern and equatorial stars. astron. Astrophys. **69**, 465

Koutchmy, S.: 1988, Space-borne coronagraphy. *Space Sci. Rev*. **47**, 95

Koutchmy, S., Lamy, P. L. 1985, Properties and Interactions of Interplanetary Dust, eds. R. H. Giese, & P. L. Lamy (Dordrecht: Reidel), 63

Lie-Svendsen, Ø., Hansteen, V. H., Leer, E., Holzer, T. E.: 2002, The Effect of Transition Region Heating on the Solar Wind from Coronal Holes. *Astrophys. J.* **566**, 562. DOI. ADS.

Menzel, D. H., Pasachoff, J. M.: 1968, On the obliteration of strong Fraunhofer lines by electron scattering in the solar corona. *Publ. Astron. Soc. Pac.* **80**, 458

Morgan, H., Cook, A. C.: 2020, The Width, Density, and Outflow of Solar Coronal Streamers. *Astrophys. J.* **893**, 57

Neckel, H., Labs, D.: 1994, Solar Limb Darkening 1986-1990 Lambda 303-nanometers to 1099-nanometers. *Solar Phys.* **153**, 91. DOI. ADS.

Reginald, N.L., Davila, J.M.: 2000, MACS For Global measurement of the Solar Wind Velocity and the Thermal Electron Temperature During the Total Solar Eclipse on 11 August 1999. *Solar Phys.* **195**, 111.

Reginald, N.L., St. Cyr, O.C., Davila, J.M., Brosius, J.W.: 2003, Electron Temperature and Speed Measurements in the Low Solar Corona: Results From the 2001 June Eclipse. *Astrophys. J.* **599**, 596.

Reginald, N. L., Davila, J. M., St. Cyr, O. C., Rabin, D. M., Guhathakurta, M., Hassler, D. M., Gashut, H.: 2011, Electron Temperatures and Flow Speeds of the Low Solar Corona: MACS Results from the Total Solar Eclipse of 29 March 2006 in Libya, *Solar Phys*. **270**, 235. DOI. ADS.

Reginald, N. L., Gopalswamy, N., Yashiro, S., Gong, Q., Guhathakurta, M.: 2017, Replacing the polarizer wheel with a polarization camera to increase the temporal resolution and reduce the overall complexity of a solar coronagraph. *J. Astron. Tel. Inst. Sys.* **3**, id. 014001

Strachan, L., Suleiman, R., Panasyuk, A. V., Biesecker, D. A., Kohl, J. L.: 2002, Empirical Densities, Kinetic Temperatures, and Outflow Velocities in the Equatorial Streamer Belt at Solar Minimum. *Astrophys. J.* **571**, 1008. DOI. ADS.

Takahashi, N., Yoneshima, W., Hiei, E.: 2000, Large-Scale Distribution of Coronal Temperature Observed at the Total Solar Eclipse on 26 February 1998. In: Last Total Solar Eclipse of the Millennium, *ASP Conference Series*, Vol. 205. Edited by W. Livingston, A. Özgüç. p. 121. ADS.

Withbroe, G. L.: 1988, The Temperature Structure, Mass, and Energy Flow in the Corona and Inner Solar Wind. *Astrophys. J.* **325**, 442. DOI. ADS.

**Supplementary Material**

**Supplement 1:** bitse_specfit_findbest.gif. Animation showing the iterative method to determine the best temperature and flow speed obtained by $\chi^2$ minimization.

**Supplement 2:** bitse_specfit_tchange.gif. Animation to determine the confidence interval of the determined temperature.

**Supplement 3:** bitse_specfit_vchange_0.97MK.gif. Animation to determine the confidence interval of the determined flow speed.